# A Bibliometric Review of Large Language Models Research from 2017 to 2023

Lizhou Fan[1]*, Lingyao Li[1], Zihui Ma[2], Sanggyu Lee[2], Huizi Yu[1], Libby Hemphill[1]

[1]School of Information, University of Michigan, Ann Arbor, MI
[2]Department of Civil and Environmental Engineering, University of Maryland, College Park, MD
*Email: lizhouf@umich.edu

# Abstract

Large language models (LLMs) are a class of language models that have demonstrated outstanding performance across a range of natural language processing (NLP) tasks and have become a highly sought-after research area, because of their ability to generate human-like language and their potential to revolutionize science and technology. In this study, we conduct bibliometric and discourse analyses of scholarly literature on LLMs. Synthesizing over 5,000 publications, this paper serves as a roadmap for researchers, practitioners, and policymakers to navigate the current landscape of LLMs research. We present the research trends from 2017 to early 2023, identifying patterns in research paradigms and collaborations. We start with analyzing the core algorithm developments and NLP tasks that are fundamental in LLMs research. We then investigate the applications of LLMs in various fields and domains including medicine, engineering, social science, and humanities. Our review also reveals the dynamic, fast-paced evolution of LLMs research. Overall, this paper offers valuable insights into the current state, impact, and potential of LLMs research and its applications.

# 1. Introduction

On March 14, 2023, OpenAI announced the release of their newest version of the large language model (LLM) – GPT-4 (OpenAI, n.d.; Sanderson, 2023). This state-of-the-art LLM powers many of OpenAI's popular AI applications, including the widely used ChatGPT, and brings much attention to the research of LLMs. An LLM is a class of language models that employs neural networks with billions of parameters, trained on gigantic amounts of unlabelled text data through self-supervised learning (Y. Shen, Heacock, et al., 2023; Zhao et al., 2023). LLMs are often based on transformers, a type of neural network architecture that is designed to process sequential data. Transformers use self-attention mechanisms to compute contextual relationships between the input tokens, allowing them to effectively capture long-range dependencies and contextual information (Vaswani et al., 2017). The emergence of LLMs in 2018 has ushered in a paradigm shift in natural language processing (NLP) research, as they have demonstrated outstanding performance across a range of tasks (Devlin et al., 2018;

Radford et al., 2018). LLMs are designed to have general-purpose capabilities, which enable them to excel across a broad spectrum of NLP tasks (Wei et al., 2022), rather than being designed solely for a single NLP task, such as sentiment analysis, named entity recognition, or text classification. Typical LLMs include Bidirectional Encoder Representations from Transformers (BERT) developed by Google (Devlin et al., 2018), Generative Pre-trained Transformer (GPT) families developed by OpenAI (Eloundou et al., 2023), and Large Language Model Meta AI (LLaMa) by Meta (Meta, 2023).

Although previous scientific literature has emphasized the potential of LLMs in various NLP tasks, including specialized applications in fields such as medical and health sciences (Ding et al., 2022; Khare et al., 2021; Yu et al., 2022) and politics (Y. Hu et al., 2022; R. Liu et al., 2021), much of the current research has been limited to specific NLP tasks or applications. With the recent release of the latest and most advanced GPT model (OpenAI, n.d.; Sanderson, 2023), LLMs have become a highly sought-after research area, attracting researchers to develop state-of-the-art LLMs, e.g. LLaMa and Bard (Meta, 2023; Pichai, 2023), and to explore their capabilities, e.g. Alpaca and GPTHuggingface (Rohan Taori, Ishaan Gulrajani, Tianyi Zhang, Yann Dubois, Xuechen Li, Carlos Guestrin, Percy Liang, Tatsunori B. Hashimoto, March, 13 2023; Y. Shen, Song, et al., 2023). Consequently, a bibliometric review to examine current LLMs research has become increasingly essential. While previous research has highlighted the potential and superiority of LLMs in NLP tasks, few studies have conducted a systematic analysis of the latest trends, opportunities, and challenges within the field of LLMs.

To gain insight into the state of LLMs research, this paper presents a comprehensive overview of current studies covering the research paradigms and collaborations in their development and applications. In particular, we focus on the discourse and bibliometric aspects, including,

- Research paradigms, the themes through topic modeling and discourse analysis of LLMs, from algorithms and NLP tasks to applications, infrastructures, and critical studies;
- Research collaborations, the scholarly collaboration networks, from the international and organizational perspectives.

The significance of this paper lies in two main aspects. Firstly, it presents an up-to-date bibliometric analysis of the state-of-the-art studies in LLMs, identifying trends and patterns that deepen understanding of the topic. Secondly, by analyzing the existing literature, our paper serves as a roadmap for researchers, practitioners, and policymakers to navigate the current landscape, pinpoint knowledge gaps and research opportunities, thereby fostering innovation and advancing the field toward breakthroughs.

# 2. Background

LLMs are pre-trained language models models that use deep learning techniques to process and comprehend natural language (Y. Shen, Heacock, et al., 2023; Zhao et al., 2023). LLMs are trained and fine-tuned on vast amounts of text data, which allows them to learn patterns in unstructured sequences and build a knowledge base of language (Brown et al., 2020; Radford et al., 2019). LLMs offer outstanding advantages over conventional NLP models. In contrast to the conventional approach for NLP tasks, which involves fine-tuning models through supervised learning on small, task-specific datasets, LLMs can effectively perform a wide range of tasks with only a few prompts (Manning, 2022). By providing them with human language descriptions or several examples of the desired task, they can execute tasks for which they were not explicitly trained (Manning, 2022). Thus, LLMs require fewer resources and less training time compared to conventional models with similar performance, as they can learn more from the same amount of data (M. Chen et al., 2021).

As such, LLMs have a broad range of capabilities in performing language-related tasks, such as text generation, translation, and summarization (Ollivier et al., 2023), as well as real-world applications, such as virtual assistants, chatbots, and language translation systems. To better outline the emerging landscapes of LLMs from 2017 to early 2023, in this section, we introduce the history of their developments, followed by their current applications across fields of research.

## 2.1 History

Traditionally, NLP models such as recurrent neural networks (RNNs) and convolutional neural networks (CNNs) had difficulty capturing long-range dependencies between words in a sentence (Hochreiter & Schmidhuber, 1997). This limitation negatively affected language models' performance on NLP tasks such as machine translation, summarization, and question-answering (Sutskever et al., 2014). However, in 2017, Vaswani et al. introduced the Transformer model (Vaswani et al., 2017). The self-attention mechanism used in the model allowed it to attend to all the other tokens in the input sequence by assigning weights to each token. The ability to capture long-range dependencies and parallelizable architecture of the model made it successful in various NLP applications (Devlin et al., 2018). Since the development of the Transformer model, researchers have built on top of the Transformer, developing more advanced language models.

BERT, which stands for Bidirectional Encoder Representations from Transformers, was introduced in 2018 by Devlin et al (Devlin et al., 2018). It is a pre-training technique that utilizes deep bidirectional representations by conditioning on both left and right contexts of all layers. This allowed the pre-trained BERT model to be fine-tuned with one additional output layer, making it suitable for a wide range of tasks such as question answering and language inference. BERT's success has led to its widespread adoption and other pre-trained language models (Y. Liu et al., 2019; Z. Yang et al., 2019). However, its limitation is that the pre-training process is computationally expensive. In 2019, Alec Radford, et al. presented Generative Pre-trained Transformer 2, also known as GPT-2, which was trained on a deep neural network with 1.5 billion parameters (Radford et al., 2019). GPT-2 utilizes a transformer architecture that employs self-attention mechanisms to gather data from various locations in the input sequence. Although the model is computationally expensive to train and run, its large size enables it to understand and generate a wide range of linguistic nuances and diverse outputs. Megatron-LM is another LLM that was developed in 2019, by (Shoeybi et al., 2019). It has 8.3 billion parameters, which is significantly larger than GPT-2's 1.5 billion parameters. This size enables the model to capture and generate more complex linguistic patterns. The model features a new parallelization scheme, which enables faster training compared to other models of comparable size. However, due to its large size, Megatron-LM requires significant computational resources for both training and inference.

In 2020, the introduction of GPT-3 by OpenAI marked a significant milestone in the development of LLMs (Brown et al., 2020). GPT-3 has 175 billion parameters, which is significantly larger than any other LLMs at that time. It can generate high-quality natural language text with little to no fine-turning due to the use of advanced techniques such as a higher layer count and more diverse training data. The introduction of GPT-3 has propelled the field of natural language processing forward. Following the success of GPT-3, researchers have continued to push the boundaries of LLMs. While these recent models are out of the time range for our analysis, they are important proofs that the advancement in LLMs research is increasingly faster – the burst of these research projects is making substantial changes to not only NLP or AI research, but also In 2023, OpenAI announced the development of a new multimodal model called GPT-4, capable of processing both text and image inputs to generate textual outputs (OpenAI, 2023). As the field is highly competitive and there are potential safety concerns, the technical paper does not disclose details about the model’s architecture, hardware, dataset construction, and training method. However, its performance was evaluated on various professional and academic exams designed for humans. GPT-4 demonstrated human-level performance on most of the exams, and notably, it achieved a score in the top 10% of test takers on a simulated version of the Uniform Bar Examination (OpenAI, 2023). There are also newly released open-source LLMs, e.g. LLaMa (Meta, 2023), which are smaller in size and number of parameters but freely available to researchers.

## 2.2 Applications

The advantages of LLMs in language understanding and their ability to generalize to new tasks have resulted in increased application and ongoing development in the field of NLP. Recent research using LLMs has focused on themes such as relation extraction (Gu et al., 2021), dialogue analysis (Thoppilan et al., 2022), text summarization (H. Zhang et al., 2019), sentiment analysis (Araci, 2019), named entity recognition (Nguyen et al., 2020), and text classification (Jin et al., 2020). These research studies have demonstrated that LLMs have the potential to significantly enhance the accuracy and fluency of natural language processing tasks, thereby improving our understanding of human language (Beltagy et al., 2019; Nguyen et al., 2020). In addition, current research on downstream tasks using LLMs has focused on several directions. One direction is fine-tuning, which involves modifying an existing pre-trained language model such as changing the weights in the neural layers by training it in a supervised fashion on a specific NLP task (Jurafsky & Martin, 2023). Another direction is the prompting interactions with LLMs (Reynolds & McDonell, 2021), where the problem to be solved is formulated via “few-shot prompting” (Brown et al., 2020) or instruction tuning (Maarten Bosma, 2021), in order to enhance the LLMs’ performance on given NLP tasks.

The versatility of LLMs makes them a promising tool for diverse disciplines and research fields. Rather than training specialized supervised models for specific tasks, researchers have utilized LLMs to handle a broad range of applications across multi-disciplinary domains. In the medical field, LLMs are used to analyze electronic health records, laboratory reports, and clinical notes to provide diagnostic assistance (Rasmy et al., 2021; Tang et al., 2021) and offer treatment suggestions (Shang et al., 2019) to healthcare professionals. They are also demonstrated to have the potential to provide AI-assisted medical education (Kung et al., 2023). In engineering, LLMs are utilized to analyze large volumes of engineering documents (Qiu & Jin, 2022), generate emergency plans (X. Liu et al., 2022), and detect and classify defects in maintaining the performance of buildings (D. U. Yang et al., 2022). Similarly, LLMs are applied to analyze social media posts, survey responses, and news articles, facilitating data-driven research in research areas such as sociology (Kawashima & Yamaguchi, 2021; Mustakim et al., 2022), economics (Jagdish et al., 2022; Li et al., 2021), and politics (Y. Hu et al., 2022; Salam et al., 2020).

# 3. Data and Methods

**Figure 1** indicates our workflow in collecting scholarly literature metadata and analyzing them. We first collected bibliometric data of LLMs research literature. Then, we analyze research paradigms and collaborations using the discourse under research themes and the scholarly collaboration networks.

**Figure 1.** Overall data and methods workflow

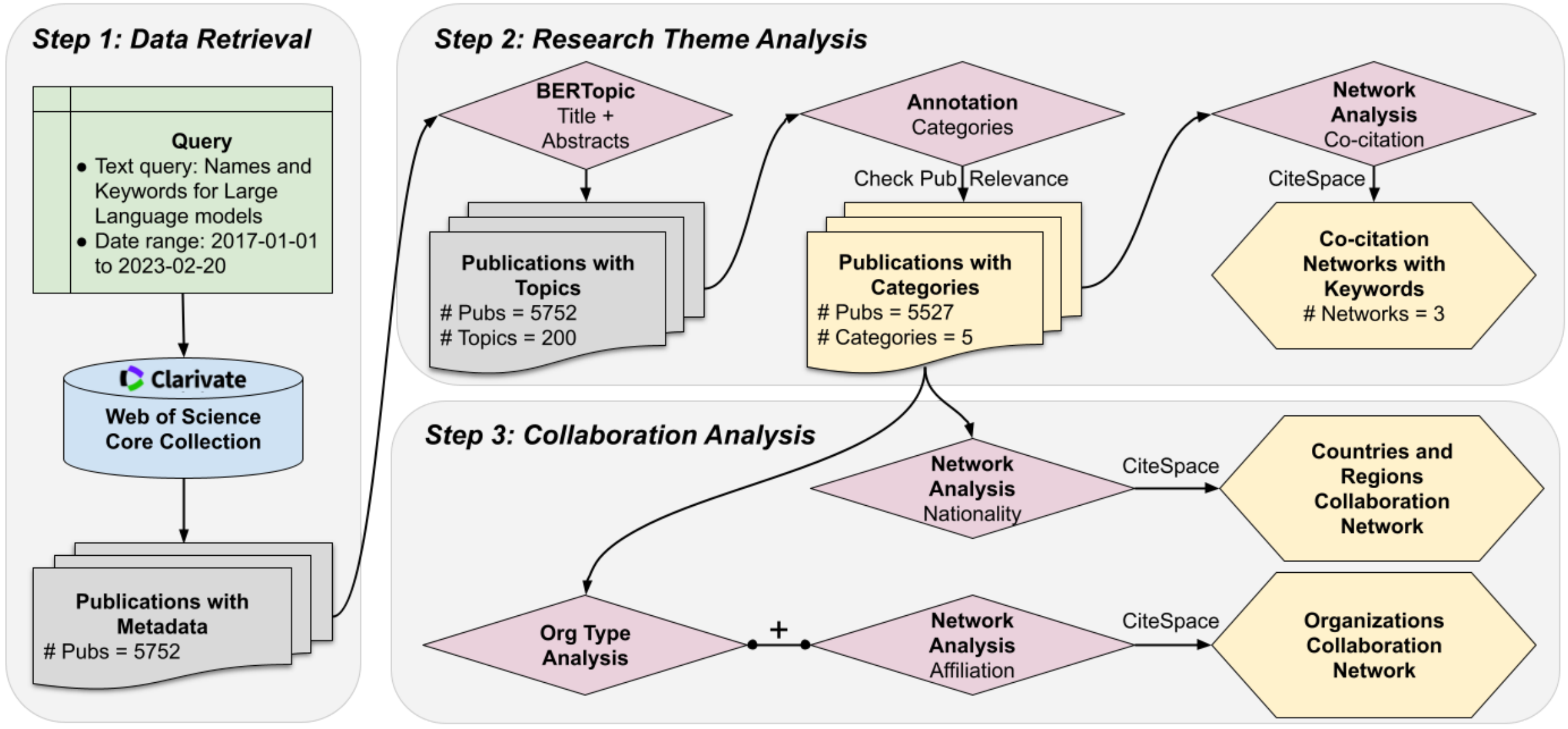

## 3.1 Data

Web of Science (WoS) Core Collection is a widely recognized platform for retrieving comprehensive academic literature metadata for bibliometric study (Birkle et al., 2020). To gather relevant papers for our analysis of LLMs research, we conducted an advanced search on WoS records. As **Table 1** shows, we then used a combination of different keywords related to the LLMs and specific models respectively on the article titles (TI) and the topics (TS), i.e. the combinations of article titles, abstracts, and keywords. The query is as follows,

```
TI=((large or big or massive) and language and (model or models)) or
TS = ("large language model" or "large language models" or BERT or
GPT-1 or GPT-2 or GPT-3 or ChatGPT)
```

We then limited the date range to the start of 2017 (2017-01-01) to early 2023 (2023-02-20) and obtained 5752 publications.

**Table 1**. Web of Science search keywords

| Search Fields | Keywords | Search Logic and Purposes |
|---|---|---|
| `TI` | large `or` big `or` massive | Combining the three components that describe possible names of LLMs using `and`. |
| | language | |
| | model `or` models | |
| `TS` | "large language model" `or` "large language models" | Combining the fixed components that describe possible names of LLMs (detectable only if consecutive) `and` popular individual LLMs names. |
| | BERT `or` GPT-1 `or` GPT-2 `or` GPT-3 `or` ChatGPT | |

## 3.2 Methods

We first used topic modeling to analyze the research paradigm of LLMs. Topic modeling is a method that discovers and summarizes latent semantic topics from large-scale unstructured text data, for example, academic literature. It assumes that each text document, for example, a publication, is a combination of multiple topics, where each topic is represented by a probability distribution of words that can be grouped as clusters with similarities (Blei & Lafferty, 2007; Steyvers, 2007).

In particular, we used BERTopic (Grootendorst, 2022), a neural topic modeling method, to analyze publications in the corpus of LLMs research. First, we use Sentence-BERT (SBERT) (Reimers & Gurevych, 2019), a transformer-based pre-trained NLP model, to obtain sentence embeddings for each of the combinations of title and abstract. In particular, we used the SBERT Python package[1] and the pre-trained model "all-MiniLM-L6-v2"[2]. To better handle the high dimensional tweet vectors for clustering, We then used a dimensionality reduction technique (UMAP) to handle the curse of dimensionality problem in our clustering model (McInnes et al., 2018), which enabled us to use the Lloyd's K-Means clustering algorithm to group similar sentences embedding vectors into topics.[3] Given the size of the corpus (more than 500 documents), we experimented with three different numbers of clusters (100, 200, and 400) and chose to use 200 clusters for our analysis since it provides publication groups that are sharing similar topics in each cluster while avoiding too many small clusters. We then proceed to represent LLMs topics and the corresponding research themes. Using the count vectorizer in the scikit-learn Python package, we tokenized topics through cluster-level (topic-level) bag-of-words representation that calculates and vectorizes the frequency of each word in each cluster (Y. Zhang et al., 2010). We then used the class-based term frequency-inverse document frequency (c-TF-IDF) to extract the difference of topical keywords (Grootendorst, 2022), distinguishing among the clusters and representing topics with the unique and frequent words as the keywords of LLMs research themes.

[1] https://www.sbert.net/
[2] https://huggingface.co/sentence-transformers/all-MiniLM-L6-v2
[3] We implemented the clustering using the scikit-learn Python package:
https://scikit-learn.org/stable/modules/generated/sklearn.cluster.KMeans.html

Finally, we characterized the 200 topics based on the keywords, titles, and abstracts of publications in each topic. After removing irrelevant contents, e.g. empty docs with "nan" and topics not about LLMs research, we summarized these topics into five higher-level categories, i.e. research themes, as follows:

- **Algorithm and NLP tasks**: The computational methods and techniques used to process, analyze, and generate human language in LLMs, performing tasks such as translation, summarization, sentiment analysis, and question-answering, among others;
- **Medical and Engineering Applications**: Applications leverage LLMs to enhance domain-specific tasks related to healthcare and engineering fields, such as analyzing medical literature, aiding diagnosis, predicting patient outcomes, and facilitating engineering design processes or problem-solving;
- **Social and Humanitarian Applications**: Applications use LLMs to address societal and humanitarian challenges, from analyzing social issues, supporting disaster response, and enhancing communication, and to promoting educational initiatives;
- **Critical Studies**: Reflections and examinations of the ethical, social, and political implications of LLMs, which scrutinize LLMs' potential biases, transparency, and impact on society, while also exploring governance, accountability, and strategies for ensuring responsible and equitable AI development and deployment;
- **Infrastructures**: The underlying systems and resources required for developing, deploying, and maintaining LLMs, examining aspects such as computing power, data storage, networking, and the policies and frameworks that govern their use and development.

To precisely identify these five research themes, as well as remove the irrelevant publications, two authors annotated all 200 topics respectively. We reached a comparatively high agreement in the first round of annotation (Krippendorff's $\alpha = 0.76$) (Krippendorff, 2018), and reached the full agreement after discussion. For the corresponding visualizations of the annotated results, as well as the LLM publications corpus, we used Tableau to create the trend line, the pie chart, and the document map (in Section 4.1).

To further study collaborations in LLMs research, we also used network methods and visualization features to study scholarly collaboration networks. In particular, we leverage a bibliometrics analysis software, CiteSpace (C. Chen, 2016), to generate co-citation and collaboration networks of LLMs publications (in Section 4.2). CiteSpace offers an essential co-citation analysis function to identify significant publications in a research field. Co-citation relationships occur when two or more papers are cited by one or more later papers at the same time. To cluster network nodes, the software employs the expectation maximization (EM) algorithm, which is an iterative algorithm that partitions data into clusters by maximizing the likelihood function based on attributes such as citation frequency and betweenness centrality (BC). The EM algorithm is a hard clustering method, which means that each reference can only belong to one cluster (C. Chen et al., 2010).

To start the clustering process, the algorithm assigns each reference to an initial random cluster, and then iteratively updates the cluster assignments based on the likelihood of the data given the cluster assignments. This process continues until the algorithm converges to a stable solution. The resultant clusters are non-overlapping and are subsequently labeled and summarized by the built-in algorithm. The co-citation knowledge graph visualizes the connections between the literature, and nodes that are closely linked in the co-citation mapping frequently appear in the same literature (Niu et al., 2022). This indicates that the co-cited articles must be similar in content, and a higher co-citation value reflects a stronger connection between them due to greater similarity in content.

Collaboration network analysis is based on social network theory, which originated from the anthropological and sociological exploration of interpersonal relationships in complex social clusters (Z. Shen et al., 2023). By analyzing the collaborative relationships between countries, institutions, and authors, CiteSpace can provide insight into the overall social status in the research field and facilitates the understanding of scholarly communication and knowledge diffusion in a particular research field. In addition, CiteSpace can track the development of a research field over time by analyzing publications from different years. This feature allows researchers to identify emerging trends and track the evolution of research areas.

# 4. Results

## 4.1 Research paradigms of LLMs: from algorithms and NLP tasks to applications, infrastructures, and critical studies

### 4.1.1 Overview of research trends and themes

The field of LLMs has gained significant attention and interest from researchers in recent years. As **Figure 2(a)** shows, there is a steady increase in the number of publications on LLMs from 2017 to 2023,[4] with a sharp spike from 2019 to 2020, likely due to increased interest in transformer-based NLP algorithms, e.g., SBERT and BERTopic (first released on September

[4] Since the total number of publications in 2023 is not yet accessible when the paper is written, we use the analytics model (additive) in Tableau to forecast the number of the publication. Based on the data from January 1, 2017 to February 20, 2023, it is predicted to have 2486 LLMs publications in 2023. We assume this forecast is conservative because of the rocketing of research interests in LLMs after the debut of ChatGPT and GPT-4 in early 2023.

2020) (Grootendorst, 2022; Reimers & Gurevych, 2019), and the public release of advanced LLM models, e.g., GPT-3 (Brown et al., 2020). The trend continues to rise in 2021 and after, indicating that the field of LLMs is still growing and evolving. This trend suggests that there is still much to explore and discover in the field of LLMs, and researchers are likely to continue studying and developing these models in the coming years.

**Figure 2.** LLMs research trends and themes

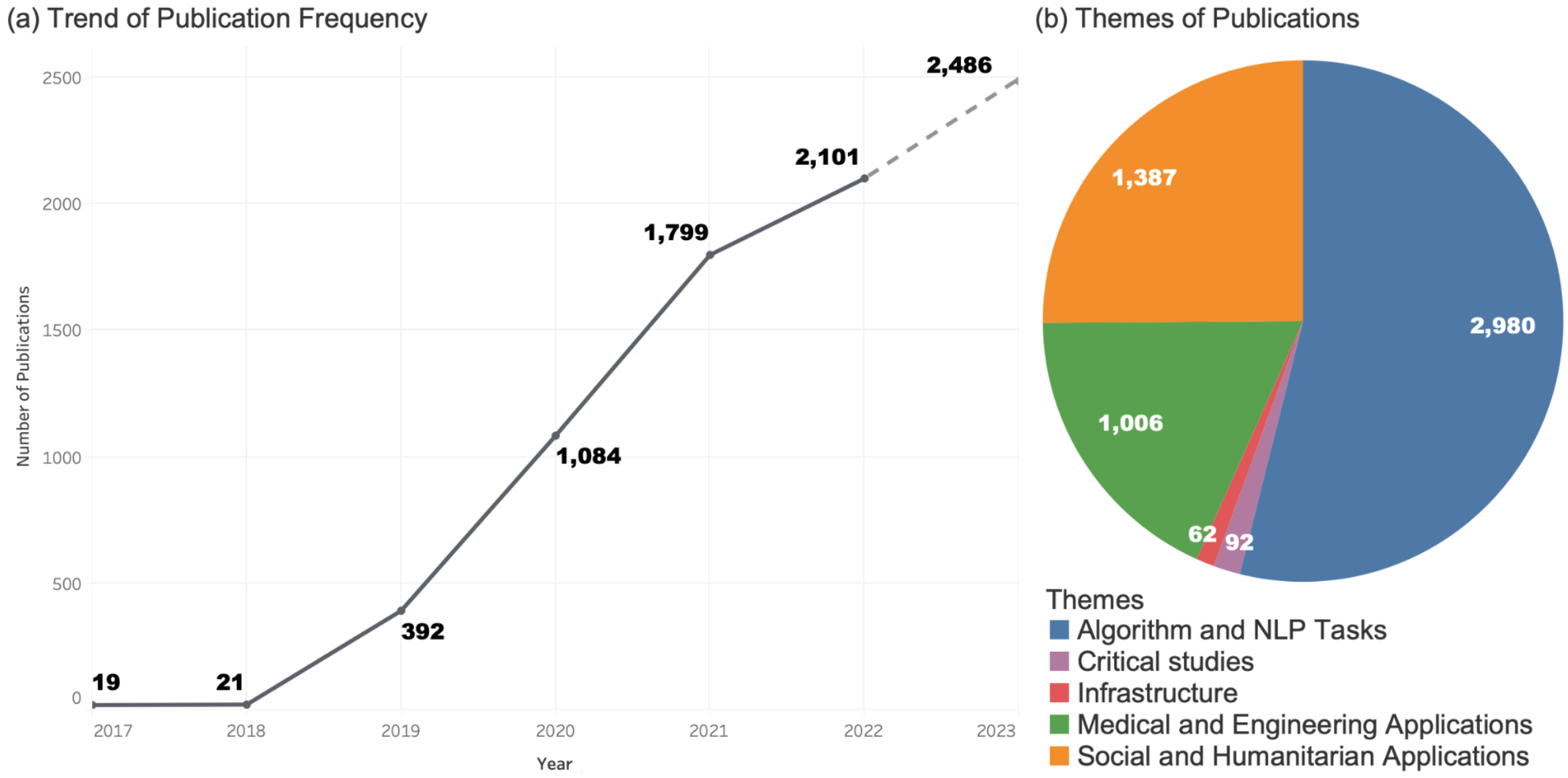

Research on LLMs also spans a wide range of themes, including Algorithm and NLP tasks, Social and Humanitarian Applications, Medical and Engineering Applications, Critical Studies, and Infrastructures. As the pie chart in **Figure 2(b)** shows, publications in the field of LLMs can be divided into several themes, each representing a specific theme or subfield. The largest research theme, Algorithm and NLP tasks, represents more than half (54%, 2980 out of 5527) of all publications in the LLM field. This theme focuses on the development and refinement of LLM architectures and modeling techniques, some of which are applicable to specific NLP tasks. The next largest research theme, Social and Humanitarian Applications, accounts for about a quarter (25%, 1387 out of 5527) of the publications. This theme includes studies that apply LLMs to specific social issues, such as controversial speech and the COVID-19 pandemic, and humanities research, such as sentiment analysis and language translation. The third largest theme, Medical and Engineering Applications, represents around 18% (1006/5527) of the publications. This theme involves the use of pre-trained LLMs and fine-tuning them to automate specific medical and engineering tasks, such as health record processing and software similarity analysis. The remaining two themes of the pie chart are relatively smaller in size, each representing less than 2% of the publications. These themes are Critical Studies, which focuses on the ethical and social implications of LLMs, and Infrastructures, which focuses on developing and enhancing hardware and cloud computing resources that can support LLMs. Overall, the snapshot of themes of publications demonstrates the different areas of focus in LLMs research, showing that the field is diverse and covers a wide range of topics and subdomains.

### 4.1.2 Topical research themes and key discourses

**Figure 3** shows a 2D mapping of topic modeling results of publications in the LLMs from 2017 to 2023.[5] Each point on the figure represents a publication.[6] This colored documentation map is divided into research theme clusters, each representing a specific group of publications that share similar topics on LLMs as defined in Section 4.1.1. In general, there are no standing-alone clusters for the larger research themes such as Algorithm and NLP Tasks and Social and Humanitarian Applications. These two themes are mixed around the map, indicating the high semantic closeness among many topics in these two clusters. For example, as the two black frames highlight, sentiment and emotion analysis are NLP tasks that require algorithm development (Topic 65 and Topic 103), which also have corresponding applications in social and humanitarian aspects (Topic 63 and Topic 28). These two themes (marked as blue and orange dots) are also scattered all around the map, which indicates that there are topics in research themes of Algorithm and NLP Tasks and Social and Humanitarian Applications themes that are related to the other three themes, demonstrating the active scholarly communication among different subdomains in LLMs research.

The map also reveals several interesting patterns in the landscape of LLMs publications, as highlighted by frames with corresponding colors. First, Medical and Engineering applications are often located on the upper middle part of the map, indicating a comprehensive cluster of various highly professional and semantically related sub-domains in LLMs research. For example, Topic 64 focuses on using LLMs to study specific categories of diseases such as Alzheimer's and Dementia, Topic 95 focuses on drug use and health services, Topic 87 focuses on biomedical advice and precision medicine, Topic 45 focuses on technique in biomedical relations extraction, and Topic 23 focuses on processing electronic medical records in different languages such as Chinese, all of which are related to medical and health research. Other subdomains in this theme, mostly engineering applications are located on other parts of the map (outside of the dotted green frame), which are closely related to the Algorithms and NLP Tasks and the Social and Humanitarian Applications research themes. Second, Critical studies (highlighted in purple dashed frames) are also semantically close to what they critically analyzed. For example, Topic 60 contains critical studies on bias in LLMs, which are closely related to LLM applications that deal with cyberbullying and abusive comments (Topic 10). Similarly, Topic 125 covers privacy concerns related to LLMs, which is closely related to anti-attack explorations in LLM applications (Topic 141). Finally, the Infrastructure theme (highlighted in red frames) focuses on parallel and distributed computing with GPU (Topic 58) and hardware and accelerator (Topic 94), which enable scalability and enhance efficiency in LLMs research.

As a whole, the map provides a visual representation of the different topics and themes that have emerged within the LLMs research community, revealing patterns and subfields that may not be immediately apparent from a simple analysis of publication keywords or titles. To further demonstrate how the topic modeling results and research themes correspond, we provide the details of the topical keywords and theme labels in **Appendix B**.

---

[5] For granular analysis, we also refer to the interactive version of this visualization on Tableau: https://public.tableau.com/app/profile/lizhou/viz/LLM_bib_categories/Documents_dash_online?publish=yes. We provide the research themes, the topic number, and the article title of each LLMs publication.

[6] The position of point (a publication) is determined by its topic distribution based on SBERT embedding in BERTopic. The x-axis represents the first principal component (PC) of the topic distribution, while the y-axis represents the second PC. Topic distributions are collections of topic word scores. Examples of topic word scores are provided in **Appendix A**.

**Figure 3.** A 2D map of LLMs publication embeddings with research themes

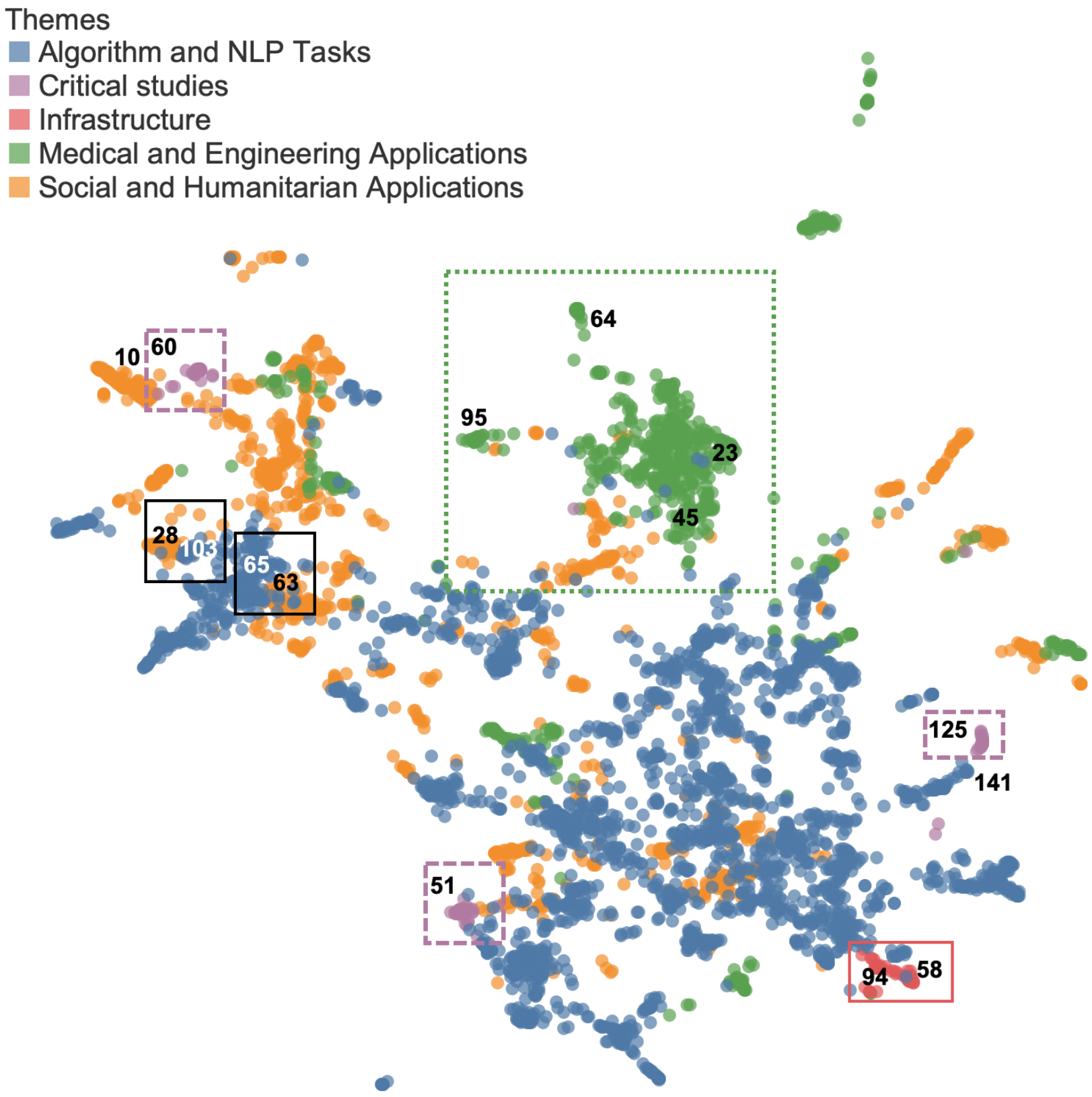

To elaborate on the key discourse under each major theme, we analyze the keywords[7] in each of the corresponding co-citation networks (**Figure 4**). In the Algorithm and NLP Tasks co-citation network (**Figure 4(a)**), the keywords of the central clusters are related to general aspects of NLP and machine learning algorithms, such as “natural language inference” (#12) and “machine learning comprehension” (#3). The peripheral clusters often have keywords with specific NLP tasks. Both the central and peripheral keywords indicate important and promising directions that have attracted attention, which are great references to new researchers and other stakeholders like publishers and funders caring about LLMs research.

In the two other co-citation networks of LLM applications, there are less obvious center clusters, which show diverse and multifaceted development among subdomains. In the Medical and Engineering Applications co-citation network (**Figure 4(b)**), the keywords suggest that the most important LLMs research themes in medical and engineering areas are related to the application of pre-trained models and NLP techniques. These applications depend on a few core NLP tasks such as named entity recognition (NER) and contextualized word embedding to support a wide range of use cases from medical tasks (e.g. clinical textual semantic similarity) to engineering

[7] Note that topical the keywords here are the Web of Science (WoS) keywords, not the topical keywords generated by the BERTopic algorithm.

(e.g. software similarity). In the Social and Humanitarian Applications co-citation network (**Figure 4(c)**), some representative keywords include “fake news”, “twitter”, “hate speech”, “rumor detection”, and “argumentation mining”. These keywords suggest that the popular themes in this sub-domain are related to the analysis of social media and news data, particularly with respect to sentiment, opinion, and controversial content.

**Figure 4.** Keywords and representative publications of major LLMs research themes

(a) Algorithm and NLP Tasks

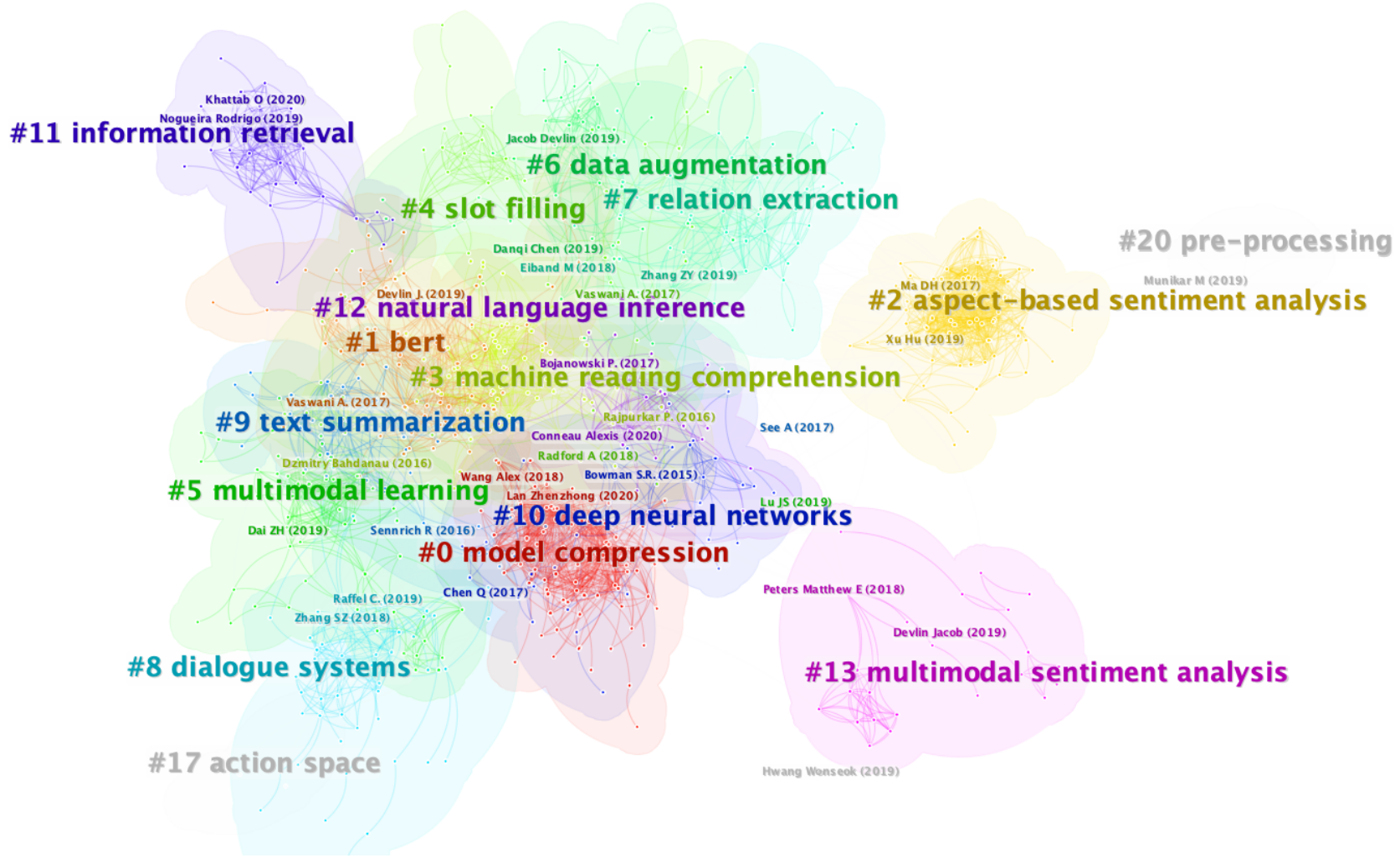

(b) Medical and Engineering Applications

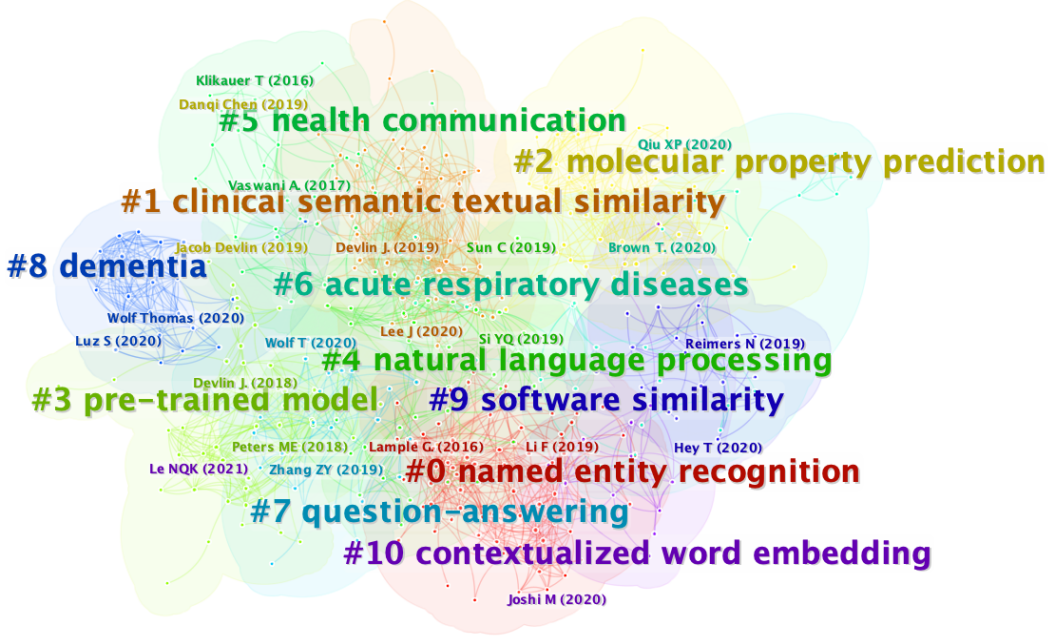

(c) Social and Humanitarian Applications

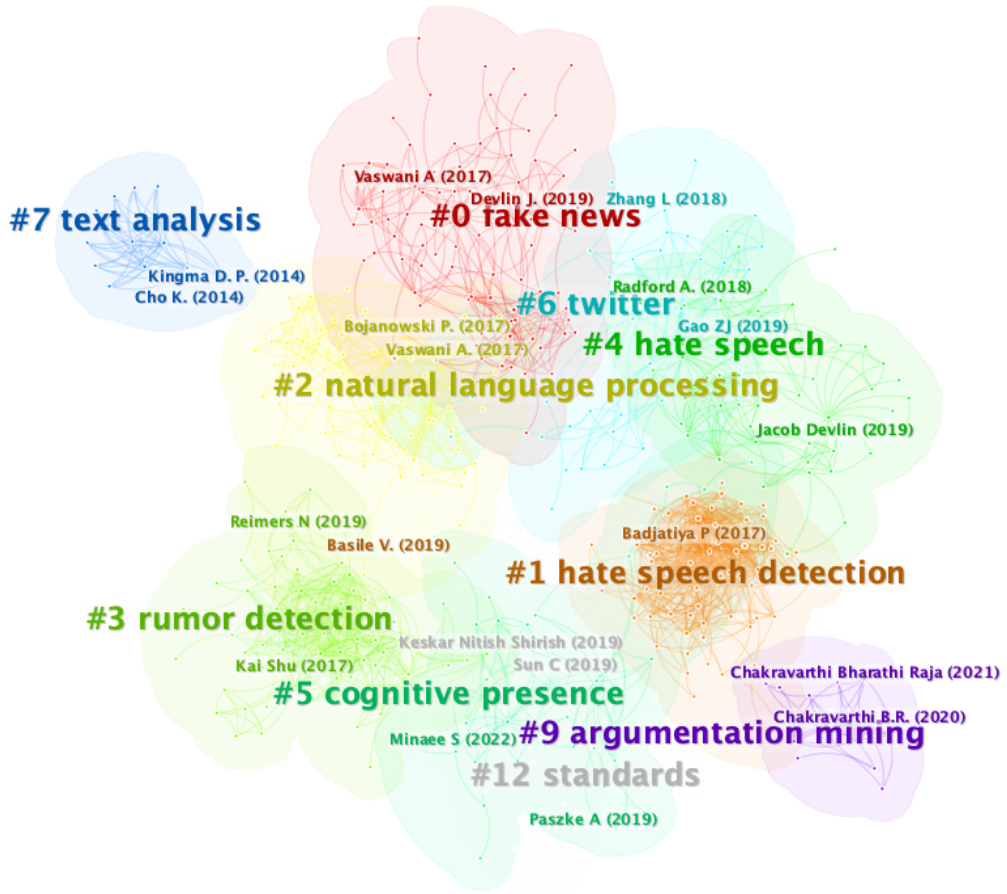

## 4.2 Research collaborations on LLMs: the international and organizational perspectives

### 4.2.1 Active countries and regions in research collaborations

**Figure 5** showcases a knowledge mapping of the distribution network for national and international collaborations, which was generated using CiteSpace. In this mapping, the centrality of a country in the collaboration network is represented by its degree, while the number of papers published from the country in our dataset is denoted by its publication frequency. Since the selected papers were sourced from recognized international journals in Web of Science, it is reasonable to conclude that the degree of centrality and frequency of publications identified in this bibliometric study reflects the importance of studies in LLM to some extent. These findings can provide valuable insights for researchers working in the field of LLM, both in current and future international collaborations.

In **Table 2**, we present the top 10 countries that have contributed the most to high-yield degree and frequency research. These countries include the United States (USA), United Kingdom (England), India, Canada, France, China, Germany, Spain, Australia, and Russia, each with influence on the collaboration network. Specifically, their respective degree values are 51, 41, 35, 34, 33, 33, 29, 28, 28, and 27. In addition, our analysis reveals that the USA, England, and India have the highest degree values of 51, 41, and 35, respectively. This suggests that these countries have the most connections with other nations in relation to LLMs research. Regarding frequency, China and the USA have the highest number of recognized publications, 1828 and 1344, respectively, greatly surpassing other countries. However, the USA has a higher degree value than China, showing a more centralized position in the collaboration network and greater outreach with other countries. Other countries and regions, such as Japan, the Netherlands, Singapore, and South Korea, are actively involved in collaboration efforts. Overall, we observed that most papers on LLMs research have been published in countries in the Asia-Pacific region, North America, and Europe.

**Figure 5**. Overall with top active countries and regions

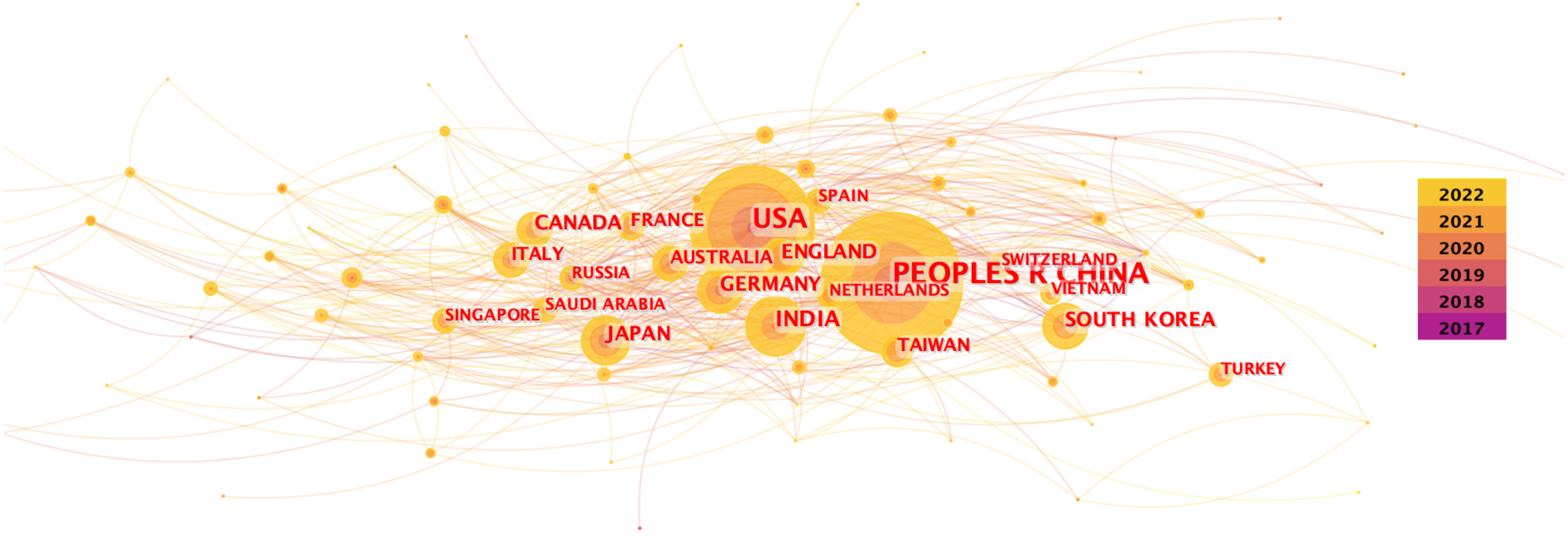

**Table 2**. Top countries and regions in the international collaboration network

| Country | Degree | Frequency |
|---|---|---|
| USA | 51 | 1344 |
| ENGLAND | 41 | 205 |
| INDIA | 35 | 377 |
| CANADA | 34 | 179 |
| FRANCE | 33 | 112 |
| PEOPLES R CHINA | 33 | 1828 |
| GERMANY | 29 | 219 |
| SPAIN | 28 | 89 |
| AUSTRALIA | 28 | 131 |
| RUSSIA | 27 | 81 |

To gain a comprehensive understanding of the collaboration process over time, we plotted a cluster analysis of international collaboration over years, as shown in **Figure 6**. The major collaborations began in 2018 among countries in different continents (**Figure 6(a)**), including USA, China, Germany, South Korea, India, England, and Japan. In 2019 (**Figure 6(b)**), additional countries and regions participated in the international collaboration related to LLMs research, including Sweden, Greece, Spain, and Italy, which were mostly from Europe.

The increasing number of participating countries in LLMs research collaborations until 2022 suggests a growing interest in this research area among researchers worldwide (**Figure 6(c)**). This trend not only highlights the popularity of LLMs research, but also demonstrates its global significance as a research domain. Overall, Figure X offers valuable insights into the evolution of international collaboration in LLMs research. Although the figure only depicts the current international collaboration status, it conveys the expansion of LLMs research beyond traditional boundaries and the emergence of new collaborative networks in this field.

**Figure 6**. Collaboration networks of active countries and regions in selected years

(a) 2018

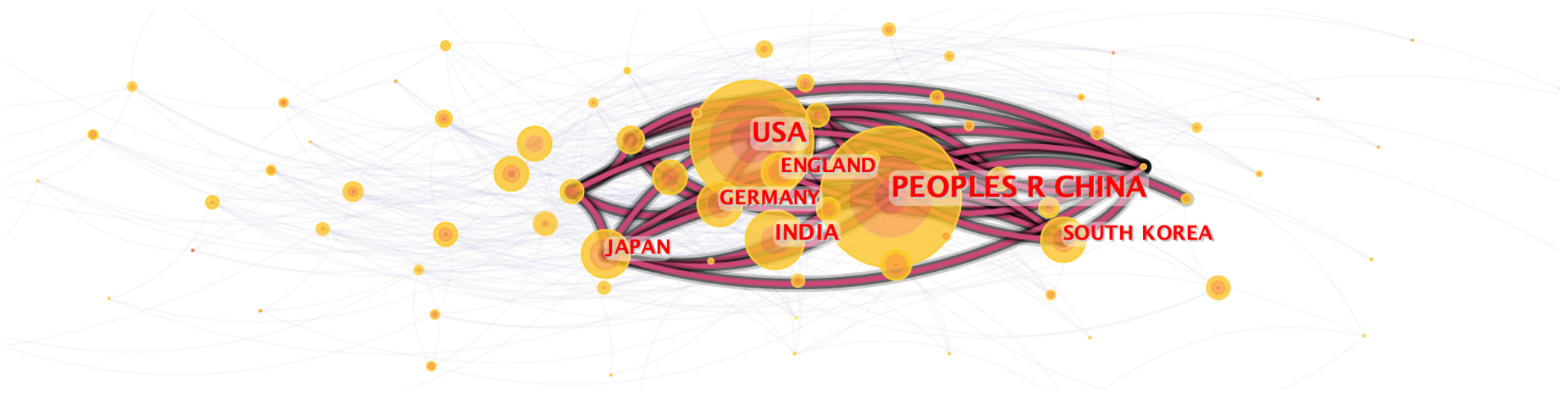

(b) 2019

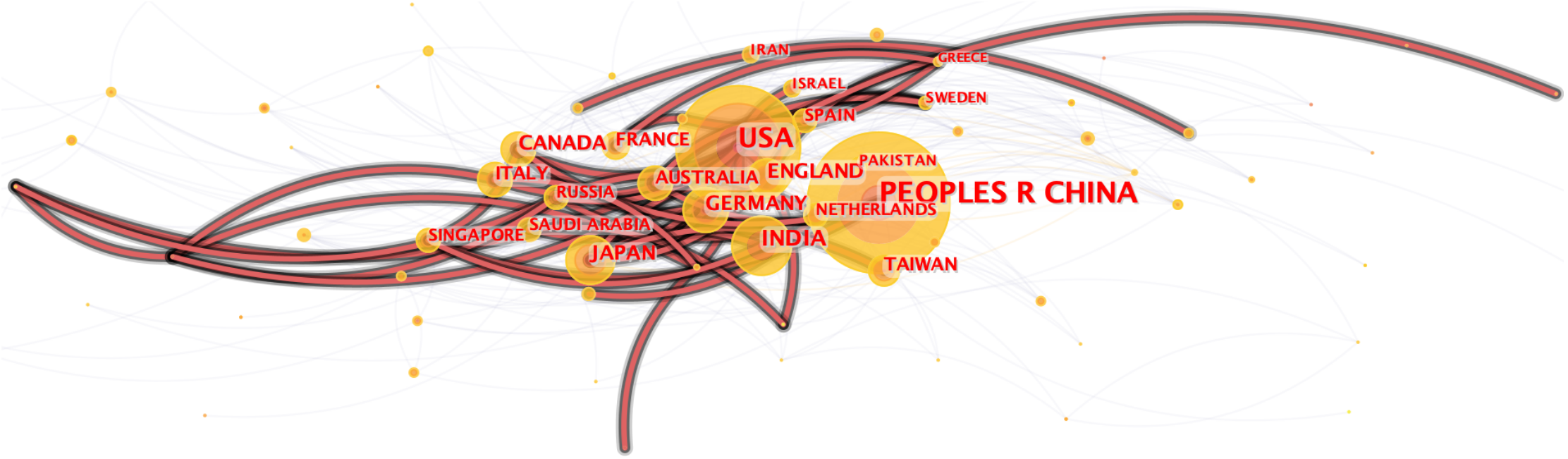

(c) 2022

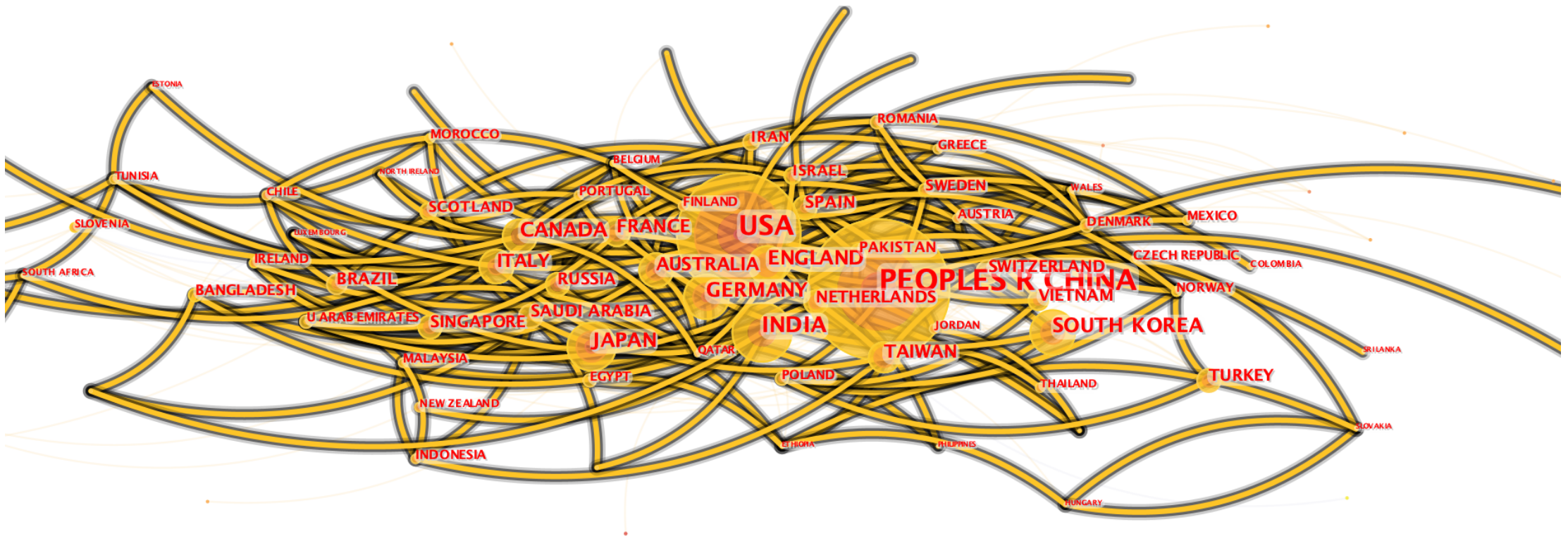

Building on the insights gained from previous analysis, we conducted further analysis and plotted the collaboration networks of active countries and regions for the identified research themes in Section 4.1. The results are presented in **Figure 7**. In examining the networks for the three most popular research themes, we found that the USA and China are at the forefront of research on algorithm and NLP tasks (**Figure 7(a)**) as well as medical and engineering applications, as illustrated by the relatively larger circle sizes in comparison to others. Additionally, we observed that several countries and regions such as Germany, England, and India remain central across all three themes, indicating their continued importance in the collaboration network.

Specifically, when examining the theme of Algorithms and NLP Tasks, we found that the USA and China are the two leading countries. However, Japan, Canada, South Korea, England, and Germany are also central in the network (**Figure 7(b)**). Russia tends to collaborate on Algorithms and Medical and Engineering Applications but does not show frequent participation in the other two research themes. In the theme of Medical and Engineering Applications, countries active in Algorithm and NLP Tasks are still active in this area. We also observed that countries or regions, such as Saudi Arabia, Australia, and Sweden, which are not actively engaged in algorithm studies, show frequent collaboration in this research theme. In comparison to the first two research themes presented in **Figure 7**, the Social and Humanitarian

Applications theme shows a more distributed network, with several countries occupying central positions (**Figure 7(c)**). For instance, India is predominantly involved in collaborations on such applications. The decentralized nature of this theme indicates that research in social and humanitarian areas is not dominated by a single country or region, but rather involves collaborations between many researchers around the world.

**Figure 7**. Collaboration networks of active countries and regions by research themes

(a) Algorithm and NLP Tasks

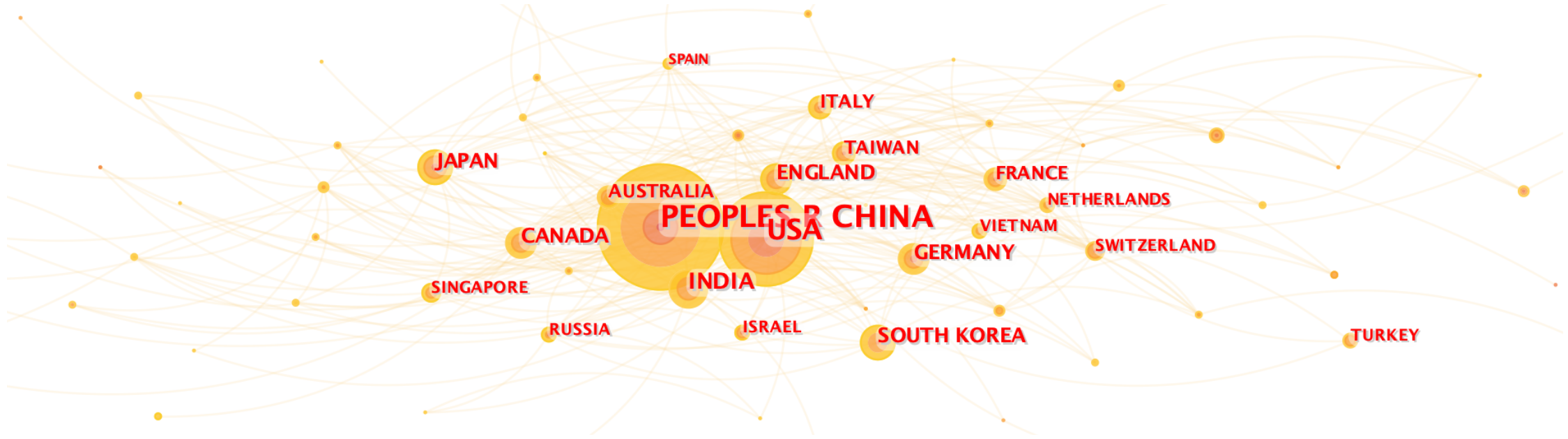

(b) Medical and Engineering Applications

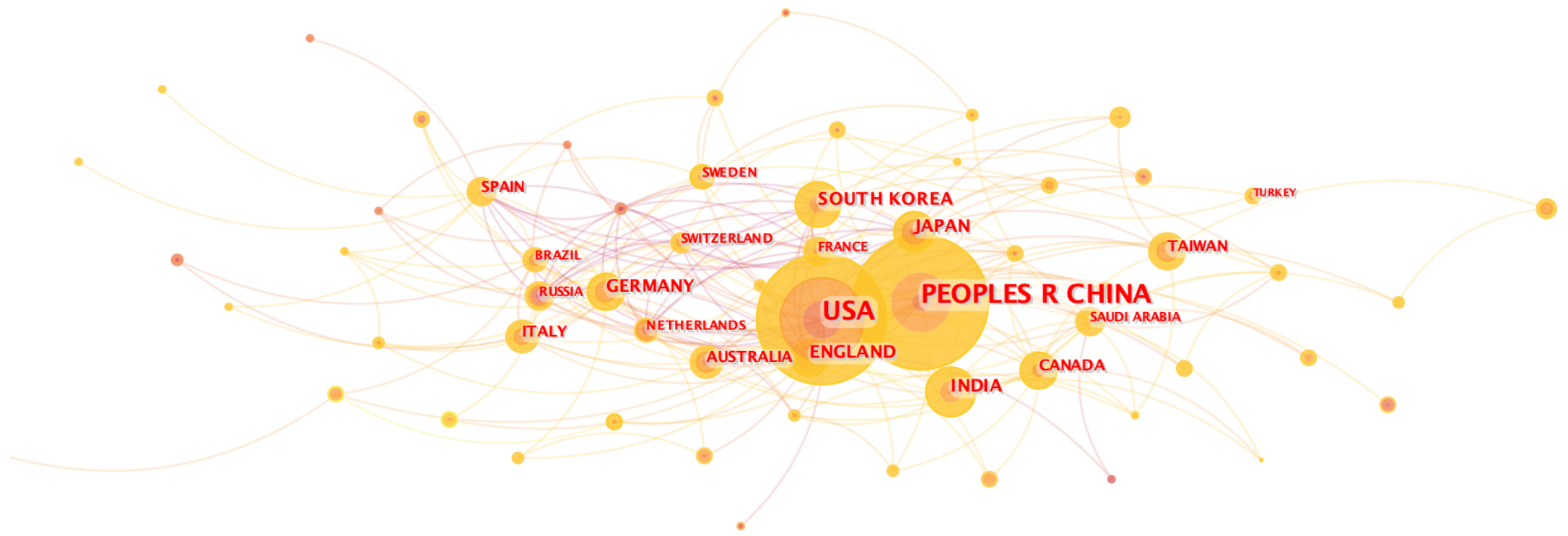

(c) Social and Humanitarian Applications

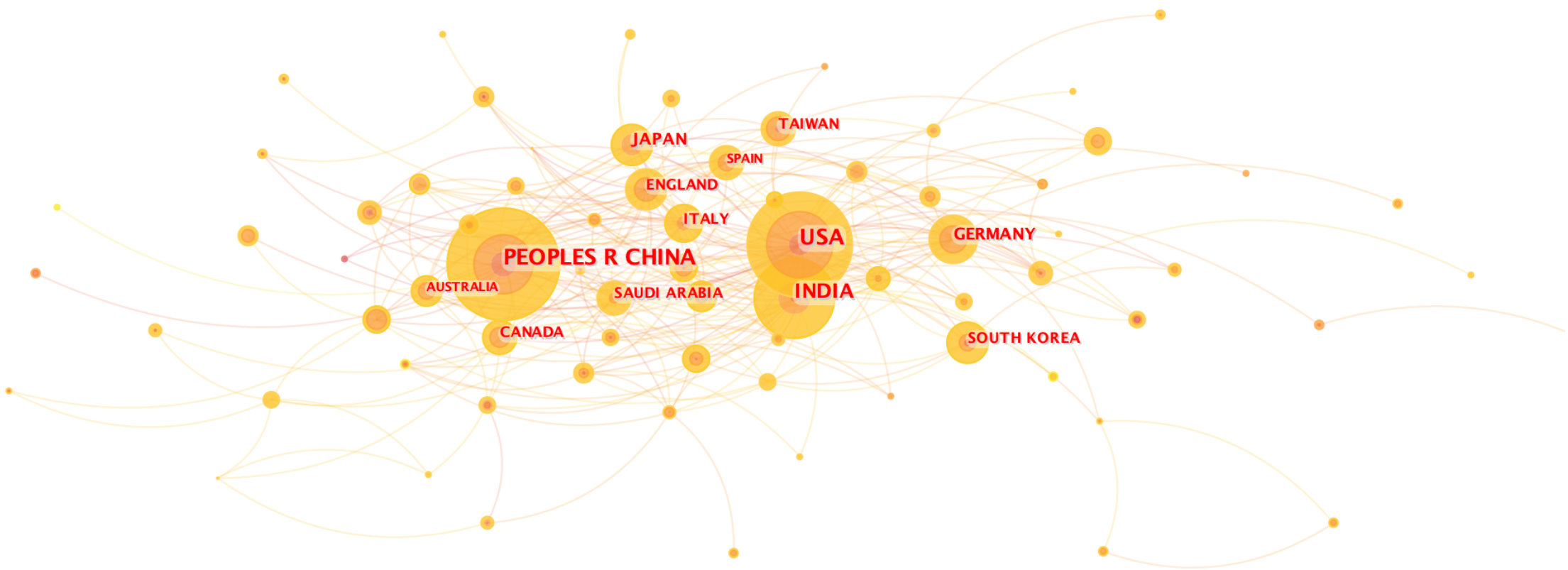

### 4.2.2 Active organizations in research collaborations

**Figure 8** displays the distribution network of contributing organizations for research papers related to LLM. **Table 3** lists the top organizations in the institutional collaboration network. The Chinese Academy of Sciences ranks first with 205 published papers, followed by the University of California System and Microsoft. Other notable institutions in the center of the network include Tsinghua University, Stanford University, Google, the University of Texas System, and Massachusetts Institute of Technology. The ranking of these institutions in terms of centrality mostly aligns with the number of papers they have published. As such, the organizations with the most published papers are also the most important and located at the center of the collaboration network. Out of the top 20 organizations with the highest degree, 14 are universities, 2 are research institutions, and 4 are tech companies.

**Figure 8** also displays the collaboration over different years. Interestingly, we observed that early works, such as publications in 2017 and 2018, are often peripheral to the network. Although these works laid the foundation for subsequent collaborations, some of the institutions that collaborated during that time did not remain at the center of the network. For instance, in 2017, New York University, the United States Navy, the United States Department of Defense, and Beijing University of Technology were involved in early works but did not stay at the center of the collaboration network. We also observed that certain institutions could carry out the implementations separately. For instance, in 2018, the Max Planck Society collaborated with other institutions to create a cluster, but its impact did not seem to endure.

Most recent research projects have been conducted through partnerships between academic and industrial organizations. Notable examples include collaborations between major tech companies and university systems, such as the joint efforts between the California University System and Stanford University with Microsoft and Google in the United States, as well as between Tsinghua University and Peking University with Tencent in China. It was also worth noting that these recent significant research projects might not have directly involved the earlier institutions mentioned. Moreover, in later years, the central entities in the collaboration network remained relatively constant, while various peripheral organizations began to contribute to the research efforts.

Regarding collaboration patterns, it was discovered that universities have been the primary contributors to LLMs research collaborations, as illustrated by that 14 out of the top 20 organizations in **Table 3** are universities. Although universities continue to play a crucial role in these endeavors, large tech corporations such as Google, Microsoft, Meta (formerly Facebook), and Tencent have also become increasingly significant collaborators. As previously noted, the combination of academic and industrial organizations has resulted in numerous significant works in this field, emphasizing the importance of cooperation between academia and industry.

**Figure 8**. Collaboration networks of active organizations, from overview and selected years

(a) Overall with top organizations

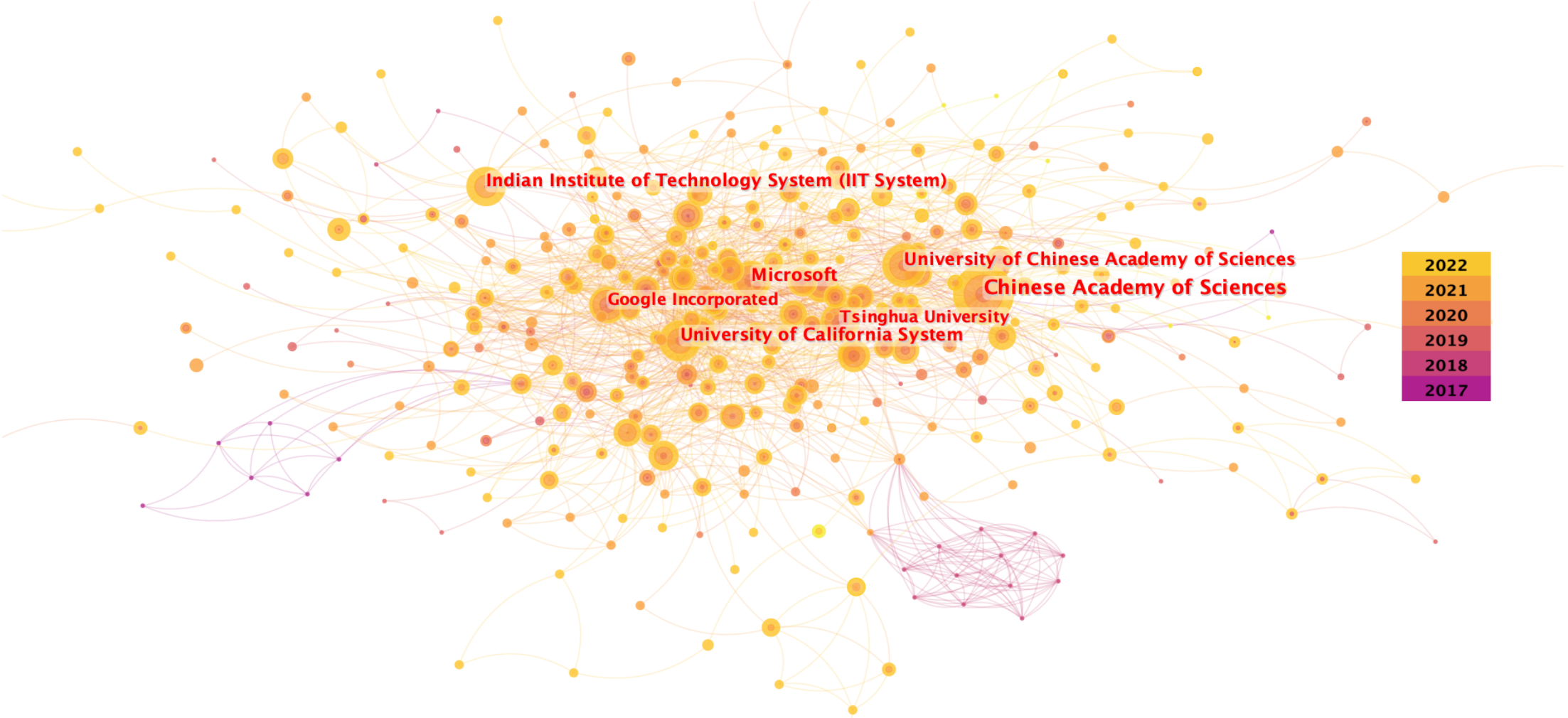

(b) 2017

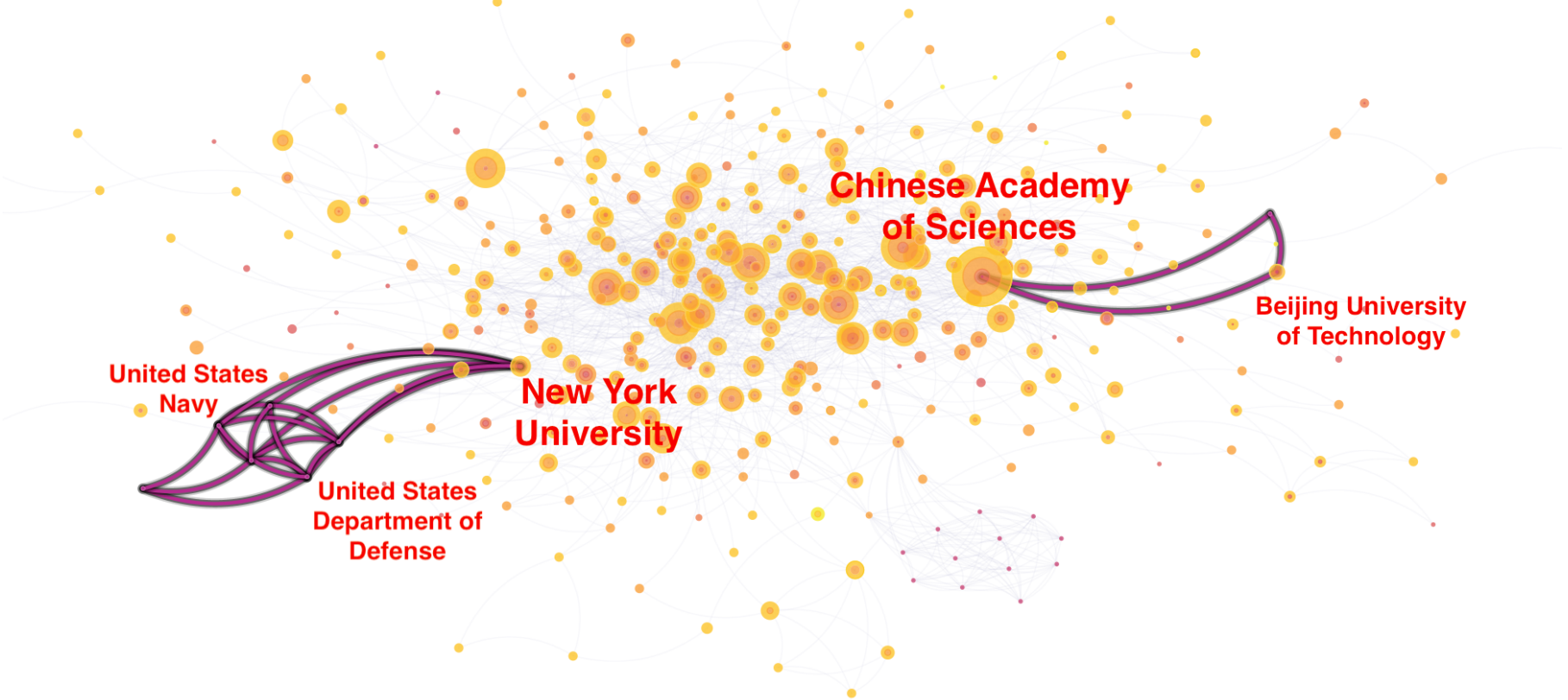

(c) 2018

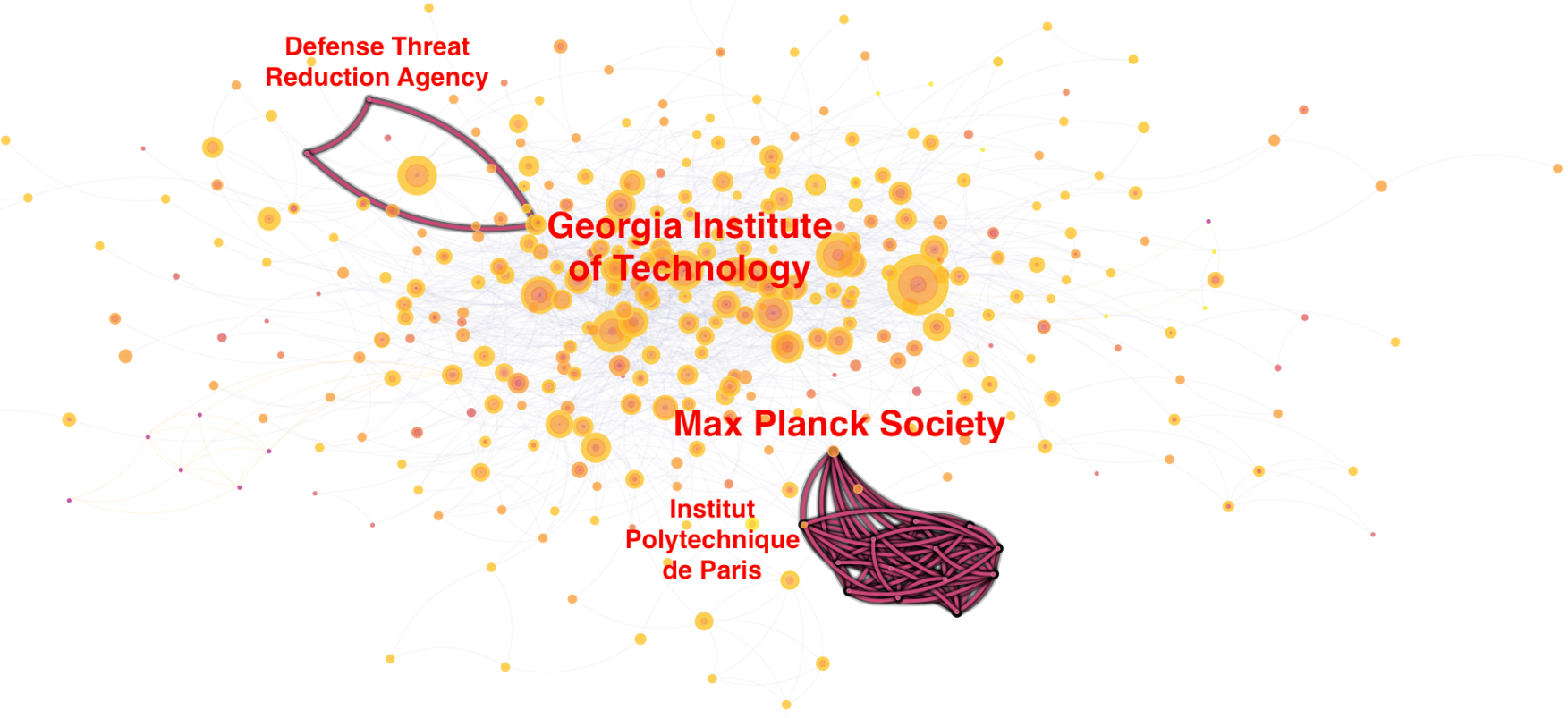

| **Table 3.** Top organizations in the institutional collaboration network | | | |
|---|---|---|---|
| **Organization** | **Frequency** | **Degree** | **Type** |
| Chinese Academy of Sciences | 205 | 60 | Research Institute |
| University of California System | 104 | 59 | University |
| Microsoft | 95 | 50 | Company |
| Tsinghua University | 86 | 43 | University |
| Stanford University | 56 | 40 | University |
| Google Incorporated | 86 | 37 | Company |
| University of Texas System | 50 | 37 | University |
| Massachusetts Institute of Technology (MIT) | 35 | 36 | University |
| University of Edinburgh | 28 | 35 | University |
| Centre National de la Recherche Scientifique (CNRS) | 60 | 35 | Research Institute |
| Tencent | 43 | 34 | Company |
| Peking University | 79 | 34 | University |
| University System of Georgia | 37 | 33 | University |
| New York University | 34 | 30 | University |
| Georgia Institute of Technology | 30 | 29 | University |
| Facebook Inc | 29 | 29 | Company |
| UDICE-French Research Universities | 53 | 28 | University |
| Pennsylvania Commonwealth System of Higher Education (PCSHE) | 27 | 28 | University |
| University of Science & Technology of China | 47 | 27 | University |
| National University of Singapore | 32 | 27 | University |

# 5. Discussion

In our discourse and bibliometric analysis, we have identified the research paradigms and collaborations of LLMs research using computational methods, namely topic modeling and network analysis. The implementation of these methods aims to provide a high-level and accurate depiction of the emerging and expanding landscapes of LLMs research.

The dynamic nature and fast evolution of LLMs research have led to significant advancements in natural language understanding and processing capabilities, with applications across diverse domains such as Medical, Engineering, Social, and Humanitarian fields. The synergistic workforce in LLMs research involving international and organizational collaborations plays a crucial role in the growth and development of this research area.

However, challenges remain due to the current movements and tensions in the development and application of LLMs. The power of LLMs is not yet clearly or openly analyzed before the release of end-user tools, such as ChatGPT and GPT-4 (Fridman, 2023). There is also a division between proponents and opponents of LLMs research and application, for instance, the open letter to pause giant AI experiments (Future of Life Institute, n.d.).

We regard our study as a glimpse of the *modern history* of LLMs research, which can be informative to newcomers of this field, policy makers of AI regulations, as well as researchers in science and technology studies. It is hard to predict the future of LLMs, while understanding its past can at least provide the knowledge foundation and warnings for future research.

## 5.1 The dynamic nature and fast evolution of LLMs research: from algorithms to applications and beyond

It is vital to recognize the dynamic nature and fast evolution, as well as the corresponding opportunities and challenges, of the research field of LLMs. The growing interest and the diverse range of themes indicate a promising future for new discoveries and advancements. As researchers continue to explore and develop novel algorithms, techniques, and applications, it is vital to recognize the dynamic nature of the research field of LLMs. The current ability of LLM algorithms has significantly improved natural language understanding and processing capabilities. These advancements have enabled researchers to tackle complex language tasks that are applicable to handle a wide range of applications across domains, including Medical, Engineering, Social, and Humanitarian fields of research. In the Social and Humanitarian domain, LLMs have been applied to analyze social media and news data, particularly with respect to sentiment, opinion, and controversial content. In the Medical and Engineering domain, LLMs are utilized to solve complex problems, from processing electronic medical records and studying specific categories of diseases to automating software similarity analysis. These diverse applications showcase the power and versatility of LLMs in addressing real-world challenges and driving advancements in various fields.

From algorithms to applications in LLMs research, there is smooth knowledge transfer among different subdomains, including specialized applications. The high semantic closeness among Algorithm and NLP Tasks and Social and Humanitarian Applications (**Figure 3**), for instance, indicates that researchers in these fields work across disciplines and share insights and expertise to develop novel solutions. This interdisciplinary approach implies that challenges in LLMs research related to Social and Humanitarian Applications are not confined to a single solution, while researchers from different backgrounds can contribute to and benefit from shared knowledge and expertise. Such collaboration can lead to more efficient problem-solving approaches and result in more impactful and far-reaching social and humanitarian applications. Moreover, the Medical and Engineering Applications theme is a comprehensive cluster of various highly professional and semantically related sub-domains in LLMs research, showing the adaptivity of LLMs algorithms. By applying pre-trained LLMs and fine-tuning them for specific tasks, researchers can leverage the power of LLMs to improve healthcare and advance engineering practices. Specifically, many top keywords among publications in Medical and Engineering Applications are general LLMs algorithms and NLP tasks, such as named entity recognition (NER) and question-answering (QA) (**Figure 4(b)**). This adaptability highlights the immense potential for LLMs to revolutionize various industries and contribute to overall societal progress.

At the same time, challenges remain in the development and application of LLMs, many of which are due the complexity and uncertainty in the dynamic nature and fast evolution of LLMs (Johanna Okerlund, Evan Klasky, Aditya Middha, Sujin Kim, Hannah Rosenfeld, Molly Kleinman, Shobita Parthasarathy, 2022; Weidinger et al., 2021). Some algorithms are not widely applied due to a variety of factors, such as computational complexity, lack of interpretability, or ethical concerns. Computational complexity can limit the scalability of certain LLMs, making them less accessible for researchers with limited resources, as well as exacerbating environmental injustice and social fragmentation. Vast computing power is required to train LLMs, coming at a significant environmental cost, while these models are not outperforming more eco-friendly models in many use cases (Goetze & Abramson, 2021). In addition, the lack of interpretability in LLMs may hinder trust and adoption in critical applications, as users may be hesitant to rely on “black box” solutions. For example, in question-answering systems or chatbots, models can mimic human-like thought and behavior, like “stochastic parrots”, without fully understanding the implications of such technology (Bender et al., 2021).

Furthermore, ethical concerns regarding biases, privacy, and other unintended consequences may prevent the widespread use of certain LLMs. First, these models can carry on existing biases in society and exacerbate them through fast and low-cost applications. For example, there is persistent anti-Muslim bias in some LLMs (Abid et al., 2021). Second, it is possible to extract personally identifiable information (PII) and other sensitive information from LLMs, raising the possibility that the massive dataset use in models can result in privacy information leaks. (Carlini et al., 2020). Other malicious uses of LLMs, such as spreading disinformation or creating fake news, can also strengthen bias and lead to social factorization problems (Guembe et al., 2022; Yamin et al., 2021). Therefore, addressing these challenges is essential to ensure the responsible and inclusive development and application of LLMs in the future.

## 5.2 The synergistic workforce in LLMs research: international and organizational collaborations

The degree of centrality and frequency shown by the scholarly collaboration networks reflect the importance of collaborations among countries and institutions studying LLMs, which can serve as a guide for researchers seeking to explore and engage in relevant research activities. These findings can also inspire and inform other stakeholders, e.g. funding agents, science and technology policymakers, and non-profit organizations, to adjust their agenda for more impactful presences in LLMs research. In general, there are valuable opportunities for researchers and other stakeholders to work together, exchange ideas, and generate knowledge that can inform policy and practice in addressing LLM needs.

The growing interest and participation in LLMs research collaborations demonstrate the global significance of this research area. We have observed that the trend towards international cooperation up to 2022 in LLM studies is gaining momentum, with an increasing number of countries and institutions joining the effort. While some countries and institutions remain at the forefront of this movement, it is encouraging to see a growing inclusive and diverse research community that brings scholars together from various backgrounds. We strongly support international collaboration for applying LLM to different contexts. In particular, one advantage we have observed is that many researchers have applied LLMs to address applications focusing on linguistic and cultural differences (J. Hu & Sun, 2020; Kim et al., 2021; Le et al., 2019).

In addition, our study provides some valuable insights for researchers who seek to identify potential partners, assess the research landscape, and discover new opportunities for collaboration. In particular, we have observed that certain institutions, such as the California University system and Stanford University in the United States, as well as Tsinghua University and Peking University in China, and companies like Microsoft and Google, have high publications and degrees in the collaboration networks. We have also observed that some institutions may have special research strength, for instance, the Indian Institute of Technology System has a strong record of publications in social and humanitarian applications. In sum, these institutions possess a strong foundation of knowledge, professional researchers, and computing resources that support LLM studies. It is essential for these leading institutions to take responsibility for the development of LLMs and provide opportunities for other institutions to join LLMs research topics in the future. For institutions interested in participating in LLMs research, seeking collaboration opportunities with these leading universities or companies could obtain access to cutting-edge resources and tools. As we have seen that some late participants in LLM studies have become influential, we believe that by leveraging these collaborations, latecomers can expand their research capabilities and contribute to the advancement of LLM studies.

Moreover, our analysis of institutional collaboration networks reveals that academia and industry maintain a close relationship in the field of LLM studies. This collaboration presents significant opportunities for both parties. Industry can provide academic researchers with access to advanced computing resources, such as cloud computing and graphics processing units, as well as financial support. Meanwhile, academia can leverage these resources to explore algorithms and solutions and help industry to test and validate real-world language processing problems. We believe that this collaboration can foster knowledge sharing between academia and industry, which can help bridge the gap between academic research and industry applications. Our findings are consistent with a previous study that emphasized the importance of strengthening the public AI research sphere in university-industry interactions to ensure equitable development of AI technology (Jurowetzki et al., 2021).

Finally, to ensure successful collaboration, we believe that it is crucial for institutions and corporations involved to understand and fulfill their roles in LLMs research. For instance, government agencies such as the United States Navy and the Department of Defense in the network, play an important role in shaping science and technology policies to regulate the applications of LLM applications in real cases. Universities, as the main body of collaboration in networks, should bring a multidisciplinary perspective to explore the research frontier such as identifying new areas of inquiry and optimizing the development of LLMs. Industry companies, which have more resources than other institutions in the network, should take social responsibility when deploying LLMs and ensure adequate supervision in place to mitigate potential risks. Data creators, whether researchers or companies, should provide specific instructions and regulations for those who use their data so that data is used ethically and in ways that align with the goals of the collaboration. Infrastructure service providers need to take into account the needs of LLMs and ensure that their infrastructure system is optimized to support these needs, such as ensuring necessary computing power and storage capacity.

## 5.3 Limitations and research outlook

There are several limitations in our study due to the scope, the method, and the availability of bibliometric data. One limitation is regarding the paper selection. Through full-text query in Web of Science Core Collection, a few papers may be irrelevant to LLMs research but got included because of including similar keywords or abbreviations. For example, a paper is selected because of the inclusion of “Bert et al.”, a citation of an author whose last name happens to be “BERT”, the abbreviation of Bidirectional Encoder Representations from Transformers (Devlin et al., 2018). We removed them based on the topic modeling results and human annotations of research themes.

One other limitation is the topic modeling process. A few papers in a topic cluster don’t look the same as other papers, not the category of the topic. For example, under the Critical Studies research theme (Topic 125), several papers have words indicating critical analyses or concerns of LLMs, e.g. “malicious”, and “social connectedness”, while they actually focus on specific engineering concepts that happened to include those keywords or relevant applications. We experimented with using SciBERT (Beltagy et al., 2019), which can improve some peripheral clustering results, while the overall topical coherence is less than the default BERT model as word embeddings. We thus stick to the default BERT model “all-MiniLM-L6-v2” which generates overall informative and comprehensive word embeddings for clustering.

There is another limitation because of the availability of LLMs research on the Web of Science. First, it is important to notice that not every large language model provides timely and public access to academia. The technical details of some LLMs are not represented by their bibliometric data. For example, the training and testing of GPT-4 are finished months before its report is available (Eloundou et al., 2023; OpenAI, n.d.). Because many LLMs research is not freely open to the public, some relevant research, especially those critics and large-scale experiments from non-partner organizations that develop the LLM, have to be delayed until there is funding or resources available. Second, some research articles on LLMs are conference papers and only exist on preprint websites such as Arxiv.org. While some LLMs are not included in our publications, the publications on the WoS core collection is representative for our analysis of international and organizational collaborations.

The field of LLMs is rapidly evolving, with new research and developments emerging at a fast pace. While this paper discusses the state-of-the-art techniques, it is likely that some of these will be surpassed by more recent advancements. As such, it is important to not only address the limitations outlined in this paper but also to stay up to date with the latest developments in LLM research. With the expected growth in the number of publications related to LLMs in 2023, we anticipate a potential publication and citation burst in the coming years, and therefore, we aim to continue monitoring this trend to ensure that our research remains relevant and impactful.

# 6. Conclusion

In this study, we have applied discourse and bibliometric analysis on over 5,000 LLMs research papers from 2017 to early 2023, surveying the emerging and expanding landscapes of their paradigms and collaborations. The rapid evolution of LLMs has resulted in significant advancements in NLP, with diverse applications across domains. Interdisciplinary, inter-organizational, and international collaborations drive these developments, fostering an inclusive research community and enabling smooth knowledge sharing. However, challenges persist, such as computational complexity, lack of interpretability, and ethical concerns in designing and applying LLMs. There is a need to call for further openness and cooperation among stakeholders, including government agencies, universities, companies, data creators, and infrastructure service providers, to ensure the responsible development and application of LLMs.

# Acknowledgment

Lizhou Fan: Conceptualization, Methodology, Analysis and interpretation of data, Writing – original draft, Visualization. Lingyao Li: Methodology, Analysis and interpretation of data, Writing – original draft, Visualization. Zihui Ma: Analysis and interpretation of data, Writing – original draft, Visualization. Sanggyu Lee: Analysis and interpretation of data, Writing – original draft. Huizi Yu: Analysis and interpretation of data, Writing – original draft, Visualization. Libby Hemphill: Writing – original draft, Supervision.

# Appendices

## A. Topic word scores

**Figure 9**. Example topic word scores

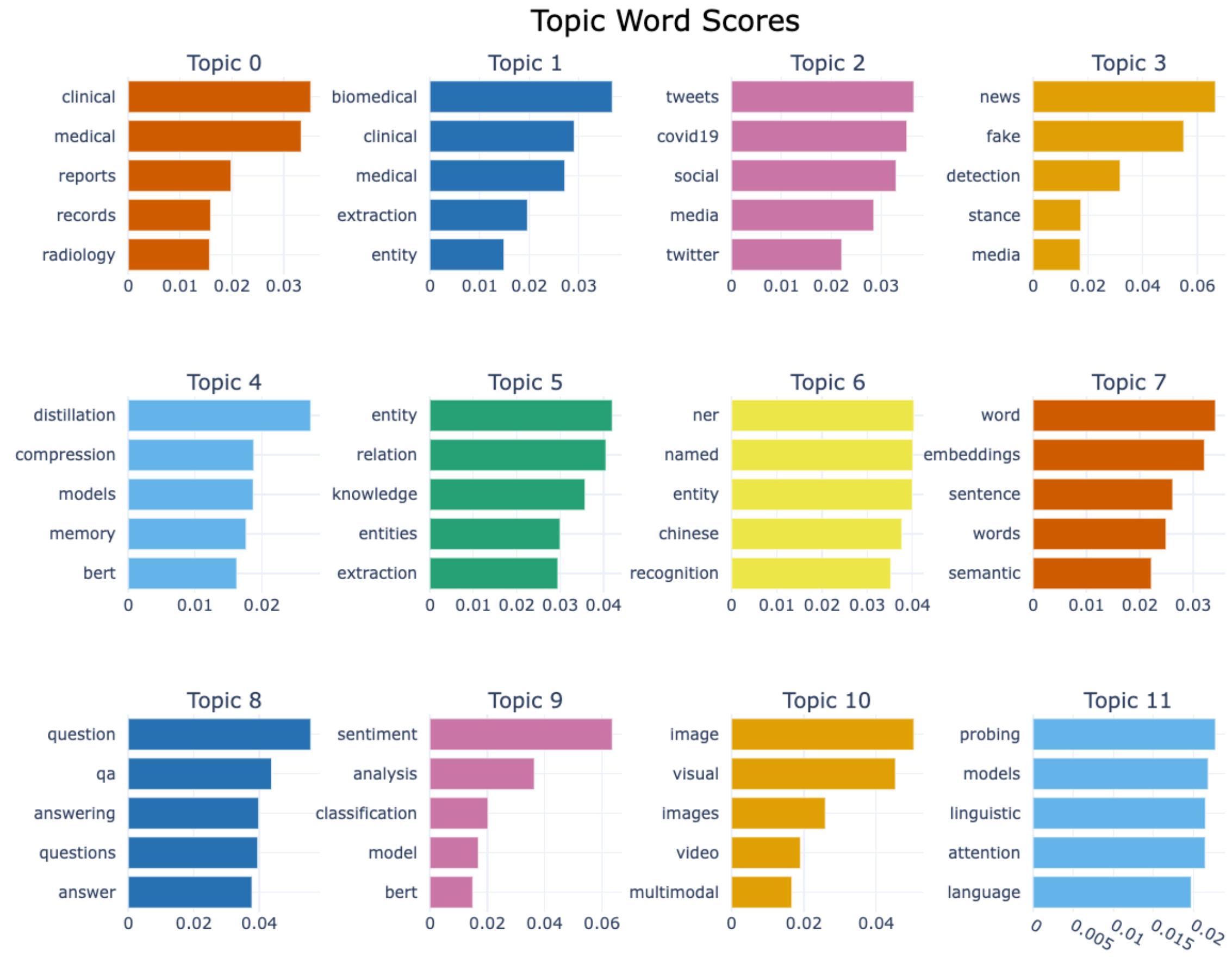

## B. Topic modeling and research themes

**Table 4**. Topics and keywords of LLMs publications

| Topic | Count | Theme | Keywords |
|---|---|---|---|
| 0 | 98 | Algorithm and NLP Tasks | 0_aspect_sentiment_aspectbased_absa |
| 1 | 98 | Algorithm and NLP Tasks | 1_ranking_retrieval_query_document |
| 2 | 89 | Algorithm and NLP Tasks | 2_visual_image_vqa_captioning |
| 3 | 82 | Medical and Engineering Applications | 3_protein_proteins_molecular_dna |
| 4 | 80 | Social and Humanitarian Applications | 4_hate_speech_offensive_hateful |
| 5 | 75 | Algorithm and NLP Tasks | 5_summarization_summary_abstractive_extractive |
| 6 | 75 | Social and Humanitarian Applications | 6_legal_law_case_judicial |
| 7 | 75 | Algorithm and NLP Tasks | 7_relation_extraction_entity_relations |
| 8 | 75 | Social and Humanitarian Applications | 8_fake_news_detection_multimodal |
| 9 | 67 | Algorithm and NLP Tasks | 9_asr_speech_spoken_acoustic |
| 10 | 66 | Social and Humanitarian Applications | 10_cyberbullying_abusive_comments_social |
| 11 | 62 | Medical and Engineering Applications | 11_disaster_disasters_events_social |
| 12 | 61 | Algorithm and NLP Tasks | 12_adversarial_attacks_attack_examples |
| 13 | 57 | Medical and Engineering Applications | 13_patients_medical_records_diagnosis |
| 14 | 57 | Medical and Engineering Applications | 14_clinical_medical_health_eligibility |
| 15 | 55 | Algorithm and NLP Tasks | 15_sentiment_analysis_migrant_classification |
| 16 | 55 | Algorithm and NLP Tasks | 16_pruning_inference_glue_compression |
| 17 | 53 | Social and Humanitarian Applications | 17_covid19_pandemic_public_tweets |
| 18 | 52 | Algorithm and NLP Tasks | 18_intent_slot_filling_intents |
| 19 | 52 | Social and Humanitarian Applications | 19_threat_vulnerability_log_cybersecurity |
| 20 | 50 | Remove | 20_nan___ |
| 21 | 50 | Social and Humanitarian Applications | 21_depression_mental_suicide_social |
| 22 | 50 | Algorithm and NLP Tasks | 22_crosslingual_multilingual_languages_mbert |
| 23 | 49 | Medical and Engineering Applications | 23_medical_chinese_electronic_records |
| 24 | 49 | Social and Humanitarian Applications | 24_reviews_review_customer_reputation |
| 25 | 48 | Algorithm and NLP Tasks | 25_translation_nmt_machine_bleu |
| 26 | 48 | Medical and Engineering Applications | 26_radiology_reports_spatial_report |
| 27 | 47 | Algorithm and NLP Tasks | 27_named_recognition_entity_chinese |
| 28 | 47 | Social and Humanitarian Applications | 28_emotion_emotions_empathy_empathetic |
| 29 | 47 | Algorithm and NLP Tasks | 29_qa_question_questions_answering |
| 30 | 46 | Algorithm and NLP Tasks | 30_answer_question_questions_reading |
| 31 | 46 | Algorithm and NLP Tasks | 31_brain_heads_attention_fmri |
| 32 | 45 | Medical and Engineering Applications | 32_correction_error_spelling_grammatical |
| 33 | 44 | Medical and Engineering Applications | 33_bug_software_code_report |
| 34 | 44 | Algorithm and NLP Tasks | 34_ner_named_entity_nested |
| 35 | 44 | Algorithm and NLP Tasks | 35_embeddings_static_words_word |
| 36 | 43 | Algorithm and NLP Tasks | 36_dialogue_persona_responses_dialog |
| 37 | 42 | Social and Humanitarian Applications | 37_stock_market_price_financial |
| 38 | 42 | Remove | 38_stevia_rebaudiana_plant_plants |
| 39 | 41 | Algorithm and NLP Tasks | 39_response_dialogue_conversation_multiturn |
| 40 | 41 | Algorithm and NLP Tasks | 40_event_causality_causal_temporal |
| 41 | 41 | Algorithm and NLP Tasks | 41_recommendation_item_sequential_recommender |
| 42 | 40 | Remove | 42_constellation_family_justice_literary |
| 43 | 39 | Algorithm and NLP Tasks | 43_knowledge_graph_graphs_entities |
| 44 | 39 | Algorithm and NLP Tasks | 44_distillation_student_teacher_kd |
| 45 | 38 | Medical and Engineering Applications | 45_relation_biomedical_extraction_relations |
| 46 | 38 | Social and Humanitarian Applications | 46_languages_morphological_multilingual_tagging |
| 47 | 38 | Algorithm and NLP Tasks | 47_linking_entity_entities_graph |
| 48 | 38 | Remove | 48_stevia_rebaudiana_antioxidant_leaves |
| 49 | 38 | Algorithm and NLP Tasks | 49_linguistic_berts_syntactic_grammar |
| 50 | 37 | Social and Humanitarian Applications | 50_citation_scientific_papers_cited |
| 51 | 37 | Critical studies | 51_ai_gpt3_intelligence_artificial |
| 52 | 37 | Social and Humanitarian Applications | 52_sarcasm_humor_irony_sarcastic |
| 53 | 37 | Algorithm and NLP Tasks | 53_emoji_emojis_sentiment_tweets |
| 54 | 37 | Algorithm and NLP Tasks | 54_clustering_topic_topics_kmeans |
| 55 | 36 | Social and Humanitarian Applications | 55_covid19_pandemic_tweets_social |
| 56 | 36 | Medical and Engineering Applications | 56_biomedical_pico_ontology_negation |
| 57 | 36 | Medical and Engineering Applications | 57_clinical_ner_medical_biomedical |
| 58 | 36 | Infrastructure | 58_memory_parallelism_gpu_distributed |
| 59 | 35 | Algorithm and NLP Tasks | 59_equipment_fault_power_knowledge |
| 60 | 35 | Critical studies | 60_bias_gender_biases_sexism |
| 61 | 35 | Algorithm and NLP Tasks | 61_chinese_segmentation_character_characters |
| 62 | 34 | Social and Humanitarian Applications | 62_scientific_science_materials_papers |
| 63 | 34 | Social and Humanitarian Applications | 63_reviews_sentiment_tourism_analysis |
| 64 | 33 | Medical and Engineering Applications | 64_ad_alzheimers_dementia_speech |
| 65 | 32 | Algorithm and NLP Tasks | 65_sentiment_analysis_software_tools |
| 66 | 32 | Medical and Engineering Applications | 66_accident_construction_defect_hazard |
| 67 | 32 | Algorithm and NLP Tasks | 67_code_codebert_program_source |
| 68 | 32 | Social and Humanitarian Applications | 68_grading_readability_students_answer |
| 69 | 32 | Algorithm and NLP Tasks | 69_probing_probe_linguistic_probes |
| 70 | 32 | Medical and Engineering Applications | 70_geant4_resistivity_proton_physics |
| 71 | 31 | Medical and Engineering Applications | 71_malware_phishing_malicious_url |
| 72 | 31 | Medical and Engineering Applications | 72_clickbait_hadith_deception_deceptive |
| 73 | 31 | Social and Humanitarian Applications | 73_idiom_words_idioms_polysemy |
| 74 | 31 | Algorithm and NLP Tasks | 74_commonsense_reasoning_knowledge_sense |
| 75 | 30 | Medical and Engineering Applications | 75_clinical_medical_notes_placebos |
| 76 | 30 | Algorithm and NLP Tasks | 76_knowledge_question_complex_answering |
| 77 | 30 | Algorithm and NLP Tasks | 77_multimodal_audio_modality_modalities |
| 78 | 30 | Algorithm and NLP Tasks | 78_comprehension_mrc_reading_answer |
| 79 | 29 | Remove | 79_species_araucaria_angustifolia_genetic |
| 80 | 29 | Algorithm and NLP Tasks | 80_sentence_similarity_sts_sentences |
| 81 | 29 | Social and Humanitarian Applications | 81_patent_patents_innovation_fintech |
| 82 | 28 | Algorithm and NLP Tasks | 82_image_multimodal_images_crossmodal |
| 83 | 28 | Algorithm and NLP Tasks | 83_qa_question_questions_answer |
| 84 | 28 | Algorithm and NLP Tasks | 84_tts_prosody_polyphone_texttospeech |
| 85 | 28 | Algorithm and NLP Tasks | 85_sentiment_sentence_representation_semantics |
| 86 | 28 | Algorithm and NLP Tasks | 86_chatbot_virtual_agent_aidir |
| 87 | 27 | Medical and Engineering Applications | 87_biomedical_advice_precision_medical |
| 88 | 27 | Medical and Engineering Applications | 88_clinical_similarity_sts_clinicalsts |
| 89 | 27 | Medical and Engineering Applications | 89_geological_geoscience_address_addresses |
| 90 | 27 | Algorithm and NLP Tasks | 90_dialogue_dst_tracking_dialog |
| 91 | 27 | Algorithm and NLP Tasks | 91_sentiment_dependency_syntactic_aspect |
| 92 | 26 | Algorithm and NLP Tasks | 92_table_tables_columns_column |
| 93 | 26 | Algorithm and NLP Tasks | 93_discourse_ordering_implicit_sentence |
| 94 | 26 | Infrastructure | 94_quantization_hardware_accelerator_rerambased |
| 95 | 26 | Medical and Engineering Applications | 95_drug_tweets_health_adverse |
| 96 | 26 | Algorithm and NLP Tasks | 96_masked_mlm_autoregressive_pretraining |
| 97 | 26 | Medical and Engineering Applications | 97_seizure_remedi_medical_care |
| 98 | 26 | Remove | 98_pt_catalysts_hydrogen_catalyst |
| 99 | 26 | Medical and Engineering Applications | 99_drug_drugs_biomedical_ds |
| 100 | 26 | Algorithm and NLP Tasks | 100_classification_text_feature_short |
| 101 | 25 | Algorithm and NLP Tasks | 101_style_gpt2_writing_dsi |
| 102 | 25 | Algorithm and NLP Tasks | 102_sign_video_sketch_image |
| 103 | 24 | Algorithm and NLP Tasks | 103_emotion_emotions_multilabel_emotional |
| 104 | 24 | Algorithm and NLP Tasks | 104_generation_revision_text_revisions |
| 105 | 24 | Social and Humanitarian Applications | 105_paraphrase_paraphrases_plagiarism_identification |
| 106 | 24 | Algorithm and NLP Tasks | 106_punctuation_disfluency_speech_asr |
| 107 | 24 | Social and Humanitarian Applications | 107_financial_finbert_finsim2_domain |
| 108 | 23 | Social and Humanitarian Applications | 108_essay_aes_scoring_essays |
| 109 | 23 | Social and Humanitarian Applications | 109_mooc_cognitive_students_moocs |
| 110 | 23 | Social and Humanitarian Applications | 110_gender_author_profiling_social |
| 111 | 23 | Algorithm and NLP Tasks | 111_adapters_adapter_token_rpt |
| 112 | 23 | Algorithm and NLP Tasks | 112_stance_argument_arguments_debate |
| 113 | 22 | Social and Humanitarian Applications | 113_hashtags_hashtag_tweet_user |
| 114 | 22 | Algorithm and NLP Tasks | 114_dependency_graph_syntactic_structure |
| 115 | 22 | Medical and Engineering Applications | 115_requirements_software_nonfunctional_requirement |
| 116 | 22 | Social and Humanitarian Applications | 116_parsing_constituency_dependency_parser |
| 117 | 21 | Social and Humanitarian Applications | 117_graph_web_page_text |
| 118 | 21 | Algorithm and NLP Tasks | 118_emotion_multimodal_audio_recognition |
| 119 | 21 | Algorithm and NLP Tasks | 119_opinion_sentiment_arabic_reviews |
| 120 | 20 | Algorithm and NLP Tasks | 120_domain_adaptation_domainspecific_domains |
| 121 | 20 | Social and Humanitarian Applications | 121_topics_public_food_safety |
| 122 | 20 | Algorithm and NLP Tasks | 122_urdu_arabic_sentiment_roman |
| 123 | 20 | Algorithm and NLP Tasks | 123_arabic_multitopic_mowjaz_arabert |
| 124 | 20 | Algorithm and NLP Tasks | 124_softmax_robustness_drift_outofdistribution |
| 125 | 20 | Critical studies | 125_privacy_traffic_sensitive_policies |
| 126 | 20 | Algorithm and NLP Tasks | 126_multilabel_labels_classification_label |
| 127 | 20 | Algorithm and NLP Tasks | 127_fewshot_prompt_gpt3_incontext |
| 128 | 19 | Algorithm and NLP Tasks | 128_document_classification_long_documents |
| 129 | 19 | Algorithm and NLP Tasks | 129_selfattention_redundancy_convolutions_finetuning |
| 130 | 19 | Algorithm and NLP Tasks | 130_chengyu_pictogram_ukrainian_bert4tc |
| 131 | 19 | Algorithm and NLP Tasks | 131_matching_siamese_retrieval_sarcnn |
| 132 | 18 | Social and Humanitarian Applications | 132_rumor_rumors_cantonese_weibo |
| 133 | 18 | Medical and Engineering Applications | 133_hpv_answering_answers_question |
| 134 | 18 | Algorithm and NLP Tasks | 134_biases_spurious_causal_visiolinguistic |
| 135 | 18 | Algorithm and NLP Tasks | 135_trajectory_traffic_missing_trajectories |
| 136 | 18 | Social and Humanitarian Applications | 136_covid19_misinformation_conspiracy_fake |
| 137 | 18 | Medical and Engineering Applications | 137_adverse_drug_adr_ade |
| 138 | 17 | Algorithm and NLP Tasks | 138_crossdomain_domain_adaptation_domains |
| 139 | 17 | Social and Humanitarian Applications | 139_personality_traits_items_psycholinguistic |
| 140 | 17 | Algorithm and NLP Tasks | 140_fewshot_label_zeroshot_labels |
| 141 | 17 | Algorithm and NLP Tasks | 141_backdoor_steganalysis_steganography_attacks |
| 142 | 17 | Algorithm and NLP Tasks | 142_dcs_transformers_cbtc_slr |
| 143 | 17 | Social and Humanitarian Applications | 143_covid19_literature_scientific_preprint |
| 144 | 16 | Social and Humanitarian Applications | 144_story_game_creative_writers |
| 145 | 16 | Social and Humanitarian Applications | 145_evidence_claim_claims_factchecking |
| 146 | 16 | Algorithm and NLP Tasks | 146_keyphrase_keyphrases_coherence_extraction |
| 147 | 16 | Algorithm and NLP Tasks | 147_federated_gradient_clients_sgd |
| 148 | 16 | Social and Humanitarian Applications | 148_product_ecommerce_attribute_products |
| 149 | 15 | Algorithm and NLP Tasks | 149_typing_entity_ultrafine_tagging |
| 150 | 15 | Social and Humanitarian Applications | 150_sports_question_tourism_answering |
| 151 | 15 | Algorithm and NLP Tasks | 151_srl_parsing_labeling_role |
| 152 | 15 | Algorithm and NLP Tasks | 152_sense_wsd_disambiguation_word |
| 153 | 15 | Medical and Engineering Applications | 153_biomedical_ner_recognition_entity |
| 154 | 14 | Social and Humanitarian Applications | 154_homes_estate_investment_smart |
| 155 | 14 | Algorithm and NLP Tasks | 155_multitask_mtdnn_jiant_gradts |
| 156 | 14 | Algorithm and NLP Tasks | 156_formality_sentence_dibert_khasi |
| 157 | 14 | Social and Humanitarian Applications | 157_spam_emails_sms_email |
| 158 | 14 | Social and Humanitarian Applications | 158_stress_tweets_twitter_buzz |
| 159 | 14 | Algorithm and NLP Tasks | 159_arithmetic_solving_problems_story |
| 160 | 13 | Social and Humanitarian Applications | 160_job_recruitment_skills_resume |
| 161 | 13 | Social and Humanitarian Applications | 161_lyrics_poetry_poems_clarity |
| 162 | 13 | Medical and Engineering Applications | 162_staff_patient_anxiety_patients |
| 163 | 13 | Algorithm and NLP Tasks | 163_lifelong_forgetting_catastrophic_update |
| 164 | 13 | Social and Humanitarian Applications | 164_movie_recommendation_book_recommender |
| 165 | 13 | Algorithm and NLP Tasks | 165_russian_complexity_outofmanifold_consert |
| 166 | 13 | Social and Humanitarian Applications | 166_political_bots_violence_republicans |
| 167 | 13 | Algorithm and NLP Tasks | 167_korean_gsl_spacing_postpositions |
| 168 | 12 | Social and Humanitarian Applications | 168_segmentation_story_paragraph_novels |
| 169 | 12 | Social and Humanitarian Applications | 169_bias_political_youtubes_news |
| 170 | 12 | Algorithm and NLP Tasks | 170_coreference_resolution_italian_mention |
| 171 | 11 | Algorithm and NLP Tasks | 171_hyperspectral_hsibert_stellar_spectra |
| 172 | 11 | Algorithm and NLP Tasks | 172_mrc_comprehension_reading_knowledge |
| 173 | 11 | Algorithm and NLP Tasks | 173_asr_speech_slu_spoken |
| 174 | 11 | Social and Humanitarian Applications | 174_service_services_web_composition |
| 175 | 11 | Algorithm and NLP Tasks | 175_amr_sequence_earlyexit_bidkt |
| 176 | 10 | Algorithm and NLP Tasks | 176_generation_action_story_stories |
| 177 | 10 | Algorithm and NLP Tasks | 177_event_enterprise_hotline_abbreviations |
| 178 | 10 | Algorithm and NLP Tasks | 178_pronoun_resolution_gendered_coreference |
| 179 | 9 | Algorithm and NLP Tasks | 179_relation_fewshot_metricbased_unseen |
| 180 | 9 | Algorithm and NLP Tasks | 180_active_al_ltp_annotation |
| 181 | 8 | Social and Humanitarian Applications | 181_email_emails_sales_transfer |
| 182 | 8 | Algorithm and NLP Tasks | 182_simplification_substitute_lexical_candidates |
| 183 | 8 | Algorithm and NLP Tasks | 183_change_lexical_shifts_contextualised |
| 184 | 8 | Social and Humanitarian Applications | 184_recipe_cooking_recipes_ingredient |
| 185 | 8 | Algorithm and NLP Tasks | 185_plms_text2chart_chart_explanations |
| 186 | 8 | Algorithm and NLP Tasks | 186_metaphor_metaphorical_vua_metaphors |
| 187 | 8 | Algorithm and NLP Tasks | 187_negation_divergences_multilingual_analogy |
| 188 | 8 | Algorithm and NLP Tasks | 188_barthez_french_lengthbased_curricula |
| 189 | 7 | Medical and Engineering Applications | 189_synthesis_program_wrangling_programming |
| 190 | 7 | Algorithm and NLP Tasks | 190_typos_hyperparameters_flitext_queries |
| 191 | 7 | Algorithm and NLP Tasks | 191_chinese_fasthan_ancient_anchibert |
| 192 | 6 | Social and Humanitarian Applications | 192_poetry_sensory_classical_va |
| 193 | 6 | Medical and Engineering Applications | 193_military_chis_identifiers_manufacturing |
| 194 | 6 | Algorithm and NLP Tasks | 194_reading_comprehension_crosslingual_czech |
| 195 | 5 | Medical and Engineering Applications | 195_covid19_pandemic_relationships_spatial |
| 196 | 5 | Algorithm and NLP Tasks | 196_kerasbert_equivalence_judging_processbert |
| 197 | 4 | Social and Humanitarian Applications | 197_music_sheet_musical_instrument |
| 198 | 4 | Algorithm and NLP Tasks | 198_coherence_utterancepair_ffcd_unilm |
| 199 | 4 | Algorithm and NLP Tasks | 199_ranking_hypotheses_nbest_reranking |